\documentclass[12pt,letterpaper]{article}
\usepackage{jheppub}
\usepackage{verbatim}
\usepackage{graphicx,times}
\usepackage{amsmath,amssymb,latexsym,float}
\usepackage{enumerate}

\numberwithin{equation}{section}

\newcommand{\abs}[1]{\left\vert#1\right\vert}

\newcommand{\R}{\text{\fontshape{n}\selectfont I\kern-.42exR}}

\newcommand{\1}{\text{\fontshape{n}\selectfont 1\kern-.56exl}}

\title{Disordered fermions, extra dimensions and a solvable Yang-Mills theory}

\author{Artan Bori\c{c}i}
\affiliation{University of Tirana\\
             Blvd. King Zog I\\
             Tirana\\
             Albania
}

\emailAdd{borici@fshn.edu.al}

\abstract{Generalizing disorder couplings of the SYK model by means of SU(N) matrices we formulate a lattice model of fermions in $d+1$ dimensions. Integration of fermions yields an effective theory of Yang-Mills fields in $d$ dimensions, the latter approaching the standard Yang-Mills theory in the case of heavy fermions and the classical limit of vanishing coupling constant of the theory. Quantum mechanically, the theory is solved using large N approximation of the dual effective theory of Hermitian matrices in $d$ dimensions. The theory is asymptotically free and confines the color. In case of massless fermions the emerging theory is an asymptotic safe QCD theory. We discuss also the relationship of this theory to the SYK model.
}

\begin{document}
\maketitle

\section{Introduction}

Yang-Mills theories and Quantum Chromodynamics (QCD) \cite{gross_wilczek,politzer,wilson,kogut_susskind} play an important role in our understanding of basic forces of the Universe. They are part of the Standard Model of particle physics. Nonetheless, an analytical solution in four dimensions is missing. Although early lattice simulations of Creutz were illuminating and showed that there is no second order phase transition between strong and weak coupling \cite{creutz}, the possibility of a Gross-Witten and Wadia transition \cite{gross_witten,wadia} diminished the hope of a full solution from the strong coupling regime. While an analytical solution is highly desirable, the high precision lattice simulation of L\"uscher and Weisz leaves no doubt that, at low energies, the Yang-Mills theory may be described by an effective string theory \cite{luscher_weisz}. Many years of research in lattice simulations of QCD have proven to be an indispensable tool in understanding the Standard Model at a non-perturbative level. During these years, string theory has made a great contribution in the search for a unifying theory of gravity and particle physics. The AdS/CFT correspondence, put forward first by Maldacena \cite{maldacena}, has established an avenue in this direction. There is recent progress made in the field using the generic model of Sachdev-Ye and Kitaev (SYK) \cite{sy,kitaev,kitaev_suh}, which uses disorder and extra dimensions as model building ingredients.

In this paper, we formulate a solvable model beyond the Standard Model. The basic degrees of freedom of the model are $d+1$ dimensional lattice fermions coupled by means of SU(N) random matrices defined on a $d$ dimensional sublattice. Such matrices generalize the randomly distributed couplings of the SYK model. However, such a generalization has a non-trivial effect in the structure of the theory: the model is chosen with the aim of obtaining a lattice gauge theory as an effective theory which remains after fermion integration. There may be an arbitrary number of such models. We have constrained the model with the intention to have a dual description of gauge theories and quantum mechanical models of holography such as SYK model and its generalizations.

Note that one may use complex $N\times N$ matrices as disorder fields. In appendix \ref{gaussian_disorder} we formulate a matrix theory with Gaussian disorder couplings. In this approach, SU(N) matrices embedded in general complex matrices are the basis of an emergent Yang-Mills theory.

In our model, we fix the length of the extra dimension by the coupling constant of the theory. Therefore, the $d$ dimensional system evolves along the extra dimension in order to reach a certain value of the coupling constant. Since each value of the latter corresponds to a given length scale, the extra dimension in our theory is the dimension of physical scales.

The results of this paper may be summarized in the following:
\begin{itemize}
\item the model with heavy fermions yields an effective theory of local Yang-Mills fields, a theory which approaches the standard Yang-Mills theory in the classical limit of vanishing coupling constant;
\item the model with massless fermions yields an effective theory of QCD with a large number of light fermions;
\item in the large $N$ limit the theory is dual to a field theory of matrices of order $N_t$, where $N_t$ is the number of lattice sites along the extra dimension;
\item the theory is non-perturbatively solvable in the large $N$ approximation;
\item the theory with heavy fermions is asymptotically free and confines the color;
\item the theory with massless fermions is dual to an asymptotic safe QCD theory, i.e. a QCD theory which is ultraviolet complete at a non-zero value of the coupling;
\item the massless fermion theory is also chirally symmetric, a symmetry which spontaneously broken;
\item although the heavy fermion theory shares important properties with the standard Yang-Mills theory like asymptotic freedom and color confinement, the corresponding beta functions are different functions of the coupling constant;
\item the asymptotic safe QCD theory, within the leading approximation of our bosonization approach and the low energy limit, is equivalent to an ideal gas of $q=2$ SYK models.
\end{itemize}

In summary, we show that a local Yang-Mills theory is solvable in the dual formulation in the large $N$ limit. Its solution shares qualitative properties of the standard Yang-Mills theory. Moreover, we are able to link gauge theories with the SYK model of holography. 

The paper is organized as follows: in the next section we define the model and discuss the disorder average of the theory. In section \ref{ym_qcd} we show that the effective description of the theory with heavy and massless fermions. In the first case we obtain a local theory Yang-Mills fields in the weak coupling limit, whereas in the second we get QCD with a large number of light modes. Section \ref{matrix_theory} deals with the disorder averaged theory which is a field theory of matrices of order $N_t$. In section \ref{solution1} we solve the matrix field theory in the large $N$ limit and heavy fermions. Section \ref{solution2} deals with the case of massless fermions in which case an asymptotic safe QCD theory emerges. In section \ref{syk2} we discuss the relation to the q=2 SYK model. The last section closes the paper with the discussion of results and their phenomenological relevance within and beyond the Standard Model. The paper includes three appendices. In appendix \ref{gaussian_disorder} we show that the model can be generalized in the case of Gaussian disorder, where we get an identical large $N$ solution as with SU(N) disorder within our approximation of bosonization of the fermion theory. In order to make the paper self contained, appendix \ref{group_integration} serves as a starting point in group integration and one-link integrals. Finally, appendix \ref{qcd_strong} applies our bosonization approach to strong coupling QCD.

\section{The model}

In this section we define our model using Hamiltonian and field theory formalism. We begin with the description of the Hamilton operator.

\subsection{Hamilton operator}
\label{hamilton_operator}

Let $\Psi_c^{\alpha}(x),\Psi_c^{\alpha}(x)^*,c=1,2,\ldots,N$ be $N$ Dirac-fermion annihilation and creation operators for each Dirac component $\alpha=1,2,\ldots,2^{d/2}$ at each site $x=(x_1,x_2,\ldots,x_d)$ on a regular Euclidean lattice of even dimensions $d$ acting on the Hilbert space ${\cal H}$. They obey the anticommutation relations:
\begin{equation}
\left\{\Psi_c^{\alpha}(x)^*,\Psi_{c'}^{\alpha'}(x')\right\}=\delta_{cc'}\delta_{\alpha\alpha'}\delta_{xx'}\ .
\end{equation}
The lattice is finite and we assume it to be a torus with $V$ number of sites. Let also $U_{\mu}(x)$ be a SU(N) matrix at each directed link $(x,x+\hat{\mu})$ on the lattice, where $\mu=1,2,\ldots,d$. The Hamiltonian operator of the model is:
\begin{equation}
\begin{aligned}
H~~=&~~~\frac{1}{a}\left\{m\sum_{x}\Psi(x)^*\gamma_5\Psi(x)\right.\\
+&~~~~~~\left.\kappa\sum_{x,\mu}\left[\Psi(x)^*\gamma_5\gamma_\mu U_{\mu}(x)\Psi(x+\hat{\mu})+\Psi(x+\hat{\mu})^*\gamma_\mu\gamma_5 U_{\mu}(x)^*\Psi(x)\right]\right\}\ ,
\end{aligned}
\end{equation}
where $a$ is the lattice spacing, $\kappa$ a dimensionless coupling constant, $\gamma_\mu$ are the usual gamma-matrices and $m$ is the bare fermion mass. We have denoted by $\gamma_5$ the product of all gamma-matrices in order to remember that the theory in four dimensions is the phenomenologically relevant theory. In the above expression we have suppressed Dirac indices for clarity. In the following the lattice spacing is set to unity. Note also that we have chosen naive fermions on the lattice. Any local discretization would do the job. For example, the Kogut-Susskind version \cite{kogut_susskind}:\footnote{For a derivation of Kogut-Susskind fermions from naive fermions see for example the book of Rothe \cite{rothe}, pages 58-59.}
\begin{equation}\label{Hamilton_operator}
\begin{aligned}
\hat H~~=&~~m\sum_{x}\hat\Psi(x)^*\hat{\gamma}_5(x)\hat\Psi(x)\\
+&~~\kappa\sum_{x,\mu}\hat{\gamma}_5(x)\eta_\mu(x)\left[\hat\Psi(x)^*U_{\mu}(x)\hat\Psi(x+\hat{\mu})+\hat\Psi(x+\hat{\mu})^*U_{\mu}(x)^*\hat\Psi(x)\right]\ ,
\end{aligned}
\end{equation}
is simpler since fermion operators carry no Dirac indices, i.e. they obey the anticommutation relations:
\begin{equation}
\left\{\hat\Psi_c(x)^*,\hat\Psi_{c'}(x')\right\}=\delta_{cc'}\delta_{xx'}\ .
\end{equation}
Here, $\hat{\gamma}_5$ is the lattice site parity operator taking $\pm1$ values on even/odd lattice sites, i.e. $\hat\gamma_5(x)=(-1)^{x_1+\cdots+x_d}$ and $\eta_1(x)=1,\eta_\mu(x)=(-1)^{x_1+\cdots+x_{\mu-1}},\mu=2,\ldots,d$. Both, $\hat\gamma_5$ and $\eta_\mu$ are diagonal matrices on the lattice. If we define hopping matrices:
\begin{equation}\label{x_dependence}
(U_\mu)_{xy;cc'}=U_\mu(x)_{cc'}\delta_{x+\hat\mu,y}\ ,
\end{equation}
then the following commutation relations hold:
\begin{equation}
\hat\gamma_5\eta_\mu-\eta_\mu\hat\gamma_5=0\ ,~~~~~~~~\hat\gamma_5U_\mu+U_\mu\hat\gamma_5=0\ ,~~~~~~~~\eta_\mu U_\mu-U_\mu\eta_\mu=0\ .
\end{equation}
For the clarity of notations we introduce the Hermitian fermion matrix:
\begin{equation}\label{fermion_matrix}
h=m\hat{\gamma}_5+\kappa \hat{\gamma}_5\sum_\mu\eta_\mu(U_\mu-U_\mu^*)\ .
\end{equation}
This way, our model may be written in the form:
\begin{equation}
\hat H=\sum_{x,y,c,c'}\hat\Psi(x)_c^*h(x,y)_{cc'}\hat\Psi(y)_{c'}\ ,
\end{equation}
where $h(x,y)_{cc'}$ are the matrix elements of the $VN\times VN$ fermionic matrix $h$.

Note that the Hamilton operator (\ref{Hamilton_operator}) defines a dynamical system in $d$ dimensions, where $d$ is the number of dimensions of the world we live in, i.e. $d=4$. The system evolves along an extra dimension, which for the moment may be thought as a fictitious time dimension. We take this extra time to be imaginary, which means that we are studying the system at a finite temperature with partition function:
\begin{equation}
Z_F(U)=\text{Tr}_{\cal H}~e^{-N_t \hat H}\ ,
\end{equation}
where the trace is taken in the Hilbert space and $N_t$ is the length of of the extra dimension in units of the lattice spacing $a$. Using the Hilbert space trace rules we get:
\begin{equation}\label{partition_function}
Z_F(U)=\det\left(1+e^{-N_t h}\right)\ .
\end{equation}
In this paper we are interested in the large $N_t$ limit of the theory. In the next subsection we compute the effective gauge action of the theory in this limit.

\subsection{Effective gauge action}

Since $h$ is traceless and using the identity $\det A=e^{\text{Tr}\log A}$ we can write the right hand side of (\ref{partition_function}) in the form:
\begin{equation}
Z_F(U)=e^{-\frac{N_t}{2}\text{Tr~}h}\det\left(e^{\frac{N_t}{2}h}+e^{-\frac{N_t}{2}h}\right)=\det\left(e^{\frac{N_t}{2}h}+e^{-\frac{N_t}{2}h}\right)\ ,
\end{equation}
where the trace is taken in the tensor product space of the lattice sites and the SU(N) group. The matrix in the last expression is an even function of $h$, therefore we get:
\begin{equation}
Z_F(U)=\det\left(e^{\frac{N_t}{2}\sqrt{h^2}}+e^{-\frac{N_t}{2}\sqrt{h^2}}\right)\ .
\end{equation}
In the large $N_t$ limit the second exponential may be neglected if $mN_t$ is not smaller than $O(1)$.\footnote{In the massless case we will assume a small mass of the order $O(1/N_t)$ with the prefactor large enough so that we may always neglect the second exponential.} With this assumption, the effective action of the theory is:
\begin{equation}\label{effective_action_h}
S_{\text{eff}}(U)=-\frac{N_t}{2}~\text{Tr~}\sqrt{h^2}\ .
\end{equation}
Eventually, we would like to integrate gauge fields and the right hand side is not convenient for this purpose. In the next subsection we formulate the theory as a $d+1$ field theory suitable to deal with the integration of the SU(N) field.

\subsection{Fermion action}

Another way to formulate the model is by introducing Grassmann valued fermion fields $\Theta(x,t)$ and $\bar{\Theta}(x,t)$, where $t=1,2,\ldots,N_t$ label the lattice sites along the extra dimension. $N_t$ assumes integer values and the Grassmann field satisfies antiperiodic boundary conditions along the extra direction. The action of the theory is:
\begin{equation}\label{action}
\tilde{{\cal I}}=\sum_{x,t}\bar{\Theta}(x,t)\hat{\partial_t}\Theta(x,t)+\sum_{x,y,t}\bar{\Theta}(x,t)h(x,y)\Theta(y,t)\ ,
\end{equation}
where $\hat{\partial_t}$ is a lattice derivative, for example:
\begin{equation}\label{lattice_derivative}
\hat{\partial_t}(t,t')=\frac{1}{2}\left(\delta_{t+1,t'}-\delta_{t-1,t'}\right)\ .
\end{equation}
In continuum limit this formulation of the model is equivalent to the Hamilton operator formulation (\ref{Hamilton_operator}).

\subsection{Gauge invariance}
\label{gauge_invariance_subsection}

The action (\ref{action}) defines a lattice theory of a massive fermion in $d+1$ dimensions in a fixed background of SU(N) random matrices, which are defined on the links of the $d$ dimensional sublattice replicated along the extra dimension. As it stands, it allows gauge variant states to propagate in the bulk. In order to ensure a full gauge invariance one can introduce an extra SU(N) field along the extra dimension and different random SU(N) matrices at each time slice of the theory, i.e. $U_\mu(x,t),\mu=1,2,\ldots,d,t=1,2,\ldots,N_t$. Fixing the axial gauge, the gauge invariant action within the restricted class of gauge transformations of the axial gauge is given by:
\begin{equation}\label{action_GI}
{\cal I}_{GI}=\sum_{x,t}\bar{\Theta}(x,t)\hat{\partial_t}\Theta(x,t)+\sum_{x,y,t}\bar{\Theta}(x,t)h_t(x,y)\Theta(y,t)\ ,
\end{equation}
where the extra dimension SU(N) field resides at the boundary. Note also that there are $N_t$ different fermion matrices at each time slice of the lattice:
\begin{equation}\label{fermion_matrices}
h_t=m\hat{\gamma}_5+\kappa \hat{\gamma}_5\sum_\mu\eta_\mu\left[U_\mu(t)-U_\mu(t)^*\right]\ ,~~~~~~~~t=1,2,\ldots,N_t\ ,
\end{equation}
where we have suppressed space and color indices for clarity (recall equation (\ref{x_dependence})). This model differs from the model (\ref{action}) in two ways: it is gauge invariant along the extra dimension and gauge fields along this dimension are $t$ dependent. The difference between these two models should be clarified in a separate study. In the following we focus with the time independent gauge field model (\ref{action}).

\subsection{Action in canonical form}

In order to bring the action in the form of a $d+1$ Dirac theory we absorb $\hat\gamma_5$ in a new set of fields:
\begin{equation}
\psi(x,t)=\Theta(x,t)\ ,~~~~~~~~\bar{\psi}(x,t)=\bar{\Theta}(x,t)\hat{\gamma}_5\ .
\end{equation}
In terms of these fields the action of equation (\ref{action}) takes the form:
\begin{equation}\label{I_action}
\begin{aligned}
{\cal I}~~~~=&~~~~\sum_{x,t}\bar{\psi}(x,t)[m+\hat{\gamma}_5(x)\hat{\partial_t}]\psi(x,t)\\
&+\kappa\sum_{x,t,\mu}\eta_\mu(x)\left[\bar{\psi}(x,t)U_{\mu}(x)\psi(x+\hat{\mu},t)-\bar{\psi}(x+\hat{\mu},t)U_{\mu}(x)^*\psi(x,t)\right]\ .
\end{aligned}
\end{equation}
We use this action in the following.

Note that we have assumed an even number of dimensions $d$ in order to be compatible with $d=4$ dimensions of the Standard Model in the case of infinite $N_t$ limit. However, for the sake of generality one may allow $d$ to be an odd integer too. In this case the Dirac field has $(d+1)/2$ components and the field $\psi(x,t)$ may not necessary be an irreducible spinor representation. Moreover, Kogut-Susskind fermions may be formulated in any dimensions.

\subsection{Disorder average}
\label{disorder_average}

Theories with disorder such as SYK use the replica trick and compute disorder average by integrating the Boltzman kernel of the replicated theory with respect to the probability distribution of the random couplings.
In the limit of large $N$ it is customary to assume a saddle point solution of decoupled replicas. Gross and Rosenhaus as well as Kitaev and Suh discuss this issue in the appendix of a recent publication on the SYK model \cite{gross_rosenhaus,kitaev_suh}. The net result of the diagonal replica assumption is that the disorder averaged theory is, in the large $N$ limit, the disorder averaged partition function of the fermion theory. This result motivates us to promote SU(N) matrices as a randomly distributed field, in which case the partition function of the theory is:
$$
Z=\int \prod_{\mu,x}dU_\mu(x)~\prod_{x,t,a}d\bar\psi(x,t)_ad\psi(x,t)_a~e^{\cal I}\ ,
$$
where $dU_\mu(x)$ is the Haar measure of the $SU(N)$ group integration and $d\bar\psi(x,t)_a$ (as well as $d\psi(x,t)_a$) is the Grassmann integration measure. This definition allows one to identify two effective theories depending which fields are integrated first. Integration of fermion fields gives a SU(N) gauge field theory in $d$ dimensions, while the integration of SU(N) fields gives an interacting fermion theory in $d+1$ dimensions. Thus, we have in principle a dual description of the same theory from the start.

\subsection{Meaning of the extra dimension}

We have defined a quantum mechanical model in $d$ dimensions, keeping in mind that $d=4$ is the physically interesting case, where one of dimensions is the time dimension. The length of the extra dimension $N_t$ is related to the value of the coupling constant $\kappa$, as it is shown in the next section. Therefore, the $d$ dimensional system evolves along the extra dimension in order to reach a certain value of $\kappa$. Since each value of $\kappa$ corresponds to a given length scale, the evolution of the system along this dimension is the evolution to reach that scale. This way, the extra dimension is also the dimension of physical scales.

In the next section we show that the $d$ dimensional effective theories derived from the model are of the Yang-Mills and QCD type.

\section{An effective theory of local Yang-Mills fields}
\label{ym_qcd}

In this section we are interested in computing an effective theory of Yang-Mills fields. In the next subsection we proceed with the case of massive fermions.

\subsection{Massive fermions}
\label{massive_fermions}

The largest fermion mass on a lattice is proportional to $m/a$ and we fix it to be exactly $1/a$. Then, form (\ref{fermion_matrix}) we have:
\begin{equation}
h^2=1+\kappa^2h_o^2\ ,~~~~~~~~~~~~h_o=\hat{\gamma}_5\sum_\mu\eta_\mu(U_\mu-U_\mu^*)\ .
\end{equation}
This way, using (\ref{effective_action_h}), the effective action of the theory is:
\begin{equation}\label{effective_action_h2}
S_{\text{eff}}(U)=-\frac{N_t}{2}~\text{Tr~}\sqrt{\1+\kappa^2h_o^2}\ .
\end{equation}
Expanding in powers of $\kappa^2$ the right hand side we get:
\begin{equation}\label{ym1}
S_{\text{eff}}(U)=c_oN_t-c_1N_t\kappa^4\sum_{\mu\nu}\text{Tr~}U_\mu U_\nu U_\mu^*U_\nu^*+O(\kappa^6)+\text{h.c.}\ ,
\end{equation}
where $c_o$ is real, $c_1=1/4$ and we have used the fact that the product of $\eta$ matrices around the plaquette equals $-1$. If we want our theory to be the standard Yang-Mills theory in the classical limit of vanishing coupling constant, then we have to make sure that the plaquette term is the Wilson discretization of the standard Yang-Mills theory. Therefore, scaling the length of the extra dimension according to the relation:
\begin{equation}\label{N_t_kappa}
c_1N_t=\frac{1}{\kappa^5}
\end{equation}
we get an effective theory which is described by the action:
\begin{equation}\label{ym2}
S_{\text{eff}}(U)=c_oN_t-\frac{1}{\kappa}\sum_{\mu\nu}\text{Tr~}U_\mu U_\nu U_\mu^*U_\nu^*+O(\kappa)+\text{h.c.}\ .
\end{equation}
When no other loops are present other than the plaquette, the theory is the Wilson discretization of the standard Yang-Mills theory. Therefore, the above theory is a theory of Yang-Mills fields enlarged by larger Wilson loops. The theory is local if the series expansion in terms of Wilson loops converges. An upper bound of the convergence radius of the series may be found using the following standard argument: since at each lattice site there are $2d-1$ choices to construct a loop without moving backwards, the number of Wilson loops of length $n$ is bounded by $(2d-1)^n$. This way, at each order $n$ in the $\kappa$ expansion, the number of Wilson loops of length $n$ does not grow faster than $(2d-1)^n$. Hence, the series converges for $\kappa\le1/(2d-1)$ and the theory is local for vanishing $\kappa$.

Classically, as $\kappa$ goes to zero, the theory is dominated by the plaquette term while larger Wilson loops may be included in a perturbation expansion. The perturbative approach in the classical theory is important if we want to maintain the correspondence principle, in which case, our theory includes the standard Yang-Mills theory as a leading order approximation of the expansion. Quantum mechanically, we may expect that the theory is asymptotically free, since larger Wilson loops are perturbations of the plaquette result. This is indeed the case as shown in section \ref{solution1}. However, the beta function of the theory vanishes linearly with the coupling constant, which shows that the approach to the continuum limit is different from what we expect in the standard Yang-Mills theory.

Note that we have related the length of the extra dimension to the coupling constant of the gauge theory. Given $N_t$ is discrete, the relation (\ref{N_t_kappa}) tells us that $\kappa$ is discrete too. However, since we are interested in the large $N_t$ limit of the theory, we assume $\kappa$ values to be real.

\subsection{Massless fermions}
\label{massless_fermions}

In case of massless fermions the fermion matrix is $\kappa h_o$. Nonetheless, we introduce a small mass $m$ of the order $O(1/N_t)$. The square of the fermion matrix in this case is:
\begin{equation}
h^2=m^2+\kappa^2h_o^2\ .
\end{equation}
In order to discover an effective local theory we start with the action of the theory (\ref{action}). By means of Fourier transformed fields $\bar{\tilde{\Theta}}(x,\omega_k)$ and $\tilde{\Theta}(x,\omega_k)$ the action may be be written in the form:
\begin{equation}
\tilde{{\cal I}}=\sum_{x,k}\bar{\tilde{\Theta}}(x,\omega_k)i\sin\omega_k\tilde{\Theta}(x,\omega_k)+\sum_{x,y,k}\bar{\tilde{\Theta}}(x,\omega_k)h(x,y)\tilde{\Theta}(y,\omega_k)\ ,
\end{equation}
where the choice:
\begin{equation}\label{omega_k}
\omega_k=\frac{\pi}{N_t}(2k+1)\ ,~~~~~~~~k=1,2,\ldots,N_t
\end{equation}
respects the boundary condition along the extra dimension. Integrating Grassmann fields the partition function may be written as a product of the following determinants:
\begin{equation}
\tilde{Z}_F(U)=\prod_{k=1}^{N_t}\det\left(h+i\sin\omega_k\right)\ .
\end{equation}
Assuming $N_t$ is even, we get:\footnote{In case $N_t$ is odd we get an extra factor $\det h$ in the product corresponding to $\omega_k=\pi$ mode.}
\begin{equation}\label{product_form}
\tilde{Z}_F(U)=\prod_{k=1}^{N_t/2}\det\left(h^2+\sin^2\omega_k\right)\ .
\end{equation}
The effective action of the theory:
\begin{equation}
S_{\text{eff}}(U)=-\sum_{k=1}^{N_t/2}\ln\left(\frac{m^2+\sin^2\omega_k}{4\kappa^2}+\frac{1}{4}h_o^2\right)
\end{equation}
is thus a theory $N_t/2$ Kogut-Susskind fermions (modulo factor four) with masses:
\begin{equation}\label{fermion_masses}
m_k=\frac{\sqrt{m^2+\sin^2\omega_k}}{2\kappa}\ ,~~~~~~~~k=1,2,\ldots,\frac{N_t}{2}\ .
\end{equation}
Note that the theory has a large number of heavy flavors with masses of the order $1/\kappa$. For such masses one can write an effective action similar to the convergent expansion (\ref{ym2}). These contribute to the Yang-Mills part of the effective action. The rest of masses $m_k$ contribute to the fermionic part of the action. This way, we end up with a local quantum field theory, which is QCD with a large number of flavors. Since a large number of frequencies scale like $1/N_t$ a large number of fermions are light with masses of the order $\kappa^4$ (for $m\sim 1/N_t$). We expect asymptotic freedom, which is a property of QCD with a limited number of massless flavors, to be lost in this case. In section \ref{solution2} we show that the theory is asymptotic safe at a fixed value of $\kappa$ which corresponds to a large value of $N_t$.

\subsection{Synthesis}

In this section we have shown that, when fermions are integrated out, the large $N_t$ limit of our model gives effective Yang-Mills and QCD theories. As shown in the rest of the paper, the dual representation of the model is solvable in the large $N$ limit. This property gives us an analytical tool to analyze these theories in this limit. In the next section we deal with the effective theory in terms of dual degrees of freedom.

\section{A Hermitian matrix field theory}
\label{matrix_theory}

The aim of this section is to formulate the theory in terms of dual fields. We first integrate gauge fields obtaining a pure fermion theory with the action $S$ defined by the equation:
\begin{equation}\label{partition_eS}
e^{S}=\int \prod_{\mu,x}dU_\mu(x)~e^{{\cal I}}\ .
\end{equation}
Then we bosonize fermions in terms of matrix fields. We begin with the gauge field integration.

\subsection{Integration of gauge fields}
\label{integration_gauge_fields}

The partition function (\ref{partition_eS}) may be written as a product of one-link integrals:
\begin{equation}\label{one_link_integrals}
\begin{aligned}
e^{S}=&e^{\sum_{x,t,t',a}\bar{\psi}_a(x,t)\left[m\delta_{t,t'}+\hat{\gamma}_5(x)\hat{\partial}_t(t,t')\right]\psi(x,t')_a}~ \prod_{\mu,x}e^{W_\mu(x)}\ ,\\
e^{W_\mu(x)}=&\int dU_\mu(x)~e^{\kappa\sum_{a,b}\eta_\mu(x)\left[\sum_t\bar{\psi}(x,t)_aU_{\mu}(x)_{ab}\psi(x+\hat{\mu},t)_b-\sum_{t'}\bar{\psi}(x+\hat{\mu},t')_aU_{\mu}(x)^*_{ab}\psi(x,t')_b\right]}\ .
\end{aligned}
\end{equation}
Group integration rules and one link integrals have been solved long ago by different authors, see for example \cite{creutz_group_integration,bars_green,brezin_gross,brower_nauenberg} and references therein. In order to make the paper self contained we have derived one-link integrals in appendix \ref{group_integration}. The result is:
\begin{equation}\label{pure_fermion}
\begin{aligned}
&S=\sum_{x,t,t',a}\bar{\psi}_a(x,t)\left[m\delta_{t,t'}+\hat{\gamma}_5(x)\hat{\partial}_t(t,t')\right]\psi(x,t')_a\\
&~~~~+N\sum_{x,\mu,t}F\left[-\frac{\kappa^2}{N^2}\sum_{t',a,b}\bar{\psi}(x,t)_b\psi(x,t')_b\bar{\psi}(x+\hat{\mu},t')_a\psi(x+\hat{\mu},t)_a\right]\ ,
\end{aligned}
\end{equation}
where the solution $F(.)$ depends on a set of boundary conditions (see (\ref{conditions})). Since we are concerned with the behavior of the theory in the large $N_t$ (i.e. small $\kappa$) limit, we require the leading order expansion of $F(.)$ in the form:
\begin{equation}\label{leading_order_approximation}
F(\Lambda)=-\Lambda+O(\Lambda^2)
\end{equation}
where, given a lattice site $x$ and direction $\mu$, $\Lambda$ is a matrix with matrix elements:
\begin{equation}
\Lambda(t_1,t_2)=-\frac{\kappa^2}{N^2}\sum_{t',a,b}\bar{\psi}(x,t_1)_b\psi(x,t')_b\bar{\psi}(x+\hat{\mu},t')_a\psi(x+\hat{\mu},t_2)_a\ .
\end{equation}
The full result is (see (\ref{one-link_result})):
\begin{equation}\label{F_function}
F(\Lambda)=1-(1+4\Lambda)^{\frac{1}{2}}+\ln\frac{1+(1+4\Lambda)^{\frac{1}{2}}}{2}\ .
\end{equation}
In this paper we use the leading order approximation (\ref{leading_order_approximation}), i.e. the action that we use in the following is:
\begin{equation}\label{psi_action}
\begin{aligned}
&S=\sum_{x,t,t',a}\bar{\psi}_a(x,t)\left[m\delta_{t,t'}+\hat{\gamma}_5(x)\hat{\partial}_t(t,t')\right]\psi(x,t')_a\\
&~~~~+\frac{\kappa^2}{N}\sum_{x,\mu,t,t',a,b}\bar{\psi}(x,t)_b\psi(x,t')_b\bar{\psi}(x+\hat{\mu},t')_a\psi(x+\hat{\mu},t)_a\ .
\end{aligned}
\end{equation}
In order to estimate the effect of higher order terms we compute the expectation value $\langle\Lambda(t_1,t_2)\rangle$. Since $\Lambda(t_1,t_2)$ fluctuations around $\langle\Lambda(t_1,t_2)\rangle$ are small in the large $N$ limit, $\Lambda(t_1,t_2)$ itself is expected to follow the behavior of $\langle\Lambda(t_1,t_2)\rangle$ in this limit. The expectation value is computed using equation (\ref{four_point_function}):
\begin{equation}
\langle\Lambda(t_1,t_2)\rangle=-\kappa^2\sum_{t'}G(x+\hat\mu,t_2,t')G(x,t',t_1)\ ,
\end{equation}
where $G(x,t,t')$ is the two point function of the theory. As shown in section \ref{solution1}, the two-point function may be written in a Fourier basis. In this basis, $\langle\Lambda\rangle$ is diagonal:
\begin{equation}
\langle\Lambda\rangle_\omega=-\kappa^2\tilde{G}_o(\omega)^2\ ,
\end{equation}
where $\tilde{G}_o(\omega)$ is defined in equation (\ref{tilde_G_o}). The right hand side is maximum at $\omega=0$ (as well as $\pi$) and the spectral radius of $\langle\Lambda\rangle$ is $\kappa^2\tilde{G}_o(0)^2$. Therefore, for massive fermions and vanishing $\kappa$ the spectral radius of $\langle\Lambda\rangle$ is $\kappa^2$, whereas for massless fermions we find $1/2d$. This way, the leading order approximation is sensible in the vanishing $\kappa$ and large $d$ limit respectively. In appendix \ref{qcd_strong}, we show also that, in the case of strong coupling QCD, where the exact result is known, the effect of higher order terms in $\Lambda$ in (\ref{leading_order_approximation}) vanishes in the large $d$ limit. We turn next to the bosonization of the fermion theory.

\subsection{Bosonization of fermions}
\label{bosonization_approach}

Let $A$ be the matrix with matrix elements:
$$
A(x,y)=\sum_{\mu}(\delta_{x+\hat\mu,y}+\delta_{x-\hat\mu,y})\ ,
$$
which connects neighbor lattice sites $x$ and $y$. Then, the fermion action may be split in two terms:
\begin{equation}
S=\sum_{x,t,t',a}\bar{\psi}_a(x,t)\left[m\delta_{t,t'}+\hat{\gamma}_5(x)\hat{\partial}_t(t,t')\right]\psi(x,t')_a+S_1\ ,
\end{equation}
with:
\begin{equation}\label{S_1_term}
S_1=\frac{\kappa^2}{2N}\sum_{x,y,t,t',a,b}\bar{\psi}(x,t)_b\psi(x,t')_bA(x,y)\bar{\psi}(y,t')_a\psi(y,t)_a\ .
\end{equation}
We integrate fermions by decoupling the quartic term, so that we end up with a quadratic action on fermion fields. If we define $\xi(x)$ matrices with matrix elements:
\begin{equation}\label{xi}
\xi(x,t,t')=\sum_{a=1}^N\bar{\psi}(x,t)_a\psi(x,t')_a\ ,
\end{equation}
then, the $S_1$ part of the action is a quadratic form in $x,y$ variables:
\begin{equation}
S_1=\frac{\kappa^2}{2N}\sum_{t,t'}\sum_{x,y}\xi(x,t,t')A(x,y)\xi(y,t',t)\ .
\end{equation}
This action can be written in the diagonal basis of $A$ as:
\begin{equation}
S_1=\frac{\kappa^2}{2N}\sum_{t,t'}\sum_{p}\tilde{A}(p)\tilde{\xi}(p,t,t')\tilde{\xi}(p,t',t)\ ,
\end{equation}
with $p$ being a four vector and $\tilde{\xi}$ and $\tilde{A}$ representations of $\xi$ and $A$ in the new basis. The exponential function $e^{S_1}$ is a product of exponential functions. These can be represented using Gaussian integrals:
\begin{equation}\label{gaussian1}
e^{\frac{\kappa^2}{2N}\tilde{A}(p)\tilde{\xi}(p,t,t)^2}=\tilde{c}_1(p)\int_{-\infty}^{+\infty}\footnotesize{d\tilde{\Sigma}(p,t,t)}~e^{-\frac{N}{2\kappa^2}\tilde{A}(p)^{-1}\tilde{\Sigma}(p,t,t)^2+\tilde{\Sigma}(p,t,t)\tilde{\xi}(p,t,t)}\ ,
\end{equation}
\begin{equation}\label{gaussian2}
e^{\frac{\kappa^2}{N}\tilde{A}(p)\tilde{\xi}(p,t,t')\tilde{\xi}(p,t',t)}=\tilde{c}_2(p)\int_{-\infty}^{+\infty}\!\!\!\!\footnotesize{d\text{Re}\tilde{\Sigma}(p,t,t')d\text{Im}\tilde{\Sigma}(p,t,t')}~e^{-\frac{N}{\kappa^2}\tilde{A}(p)^{-1}\abs{\tilde{\Sigma}(p,t,t')}^2+\tilde{\Sigma}(p,t,t')\tilde{\xi}(p,t',t)+\overline{\tilde{\Sigma}(p,t,t')}\tilde{\xi}(p,t,t')}\ ,
\end{equation}
where $\tilde{c}_1(p),\tilde{c}_2(p)$ are integration constants. This way, one finds (modulo an integration constant):
\begin{equation}
e^{S_1}=\int\prod_{p,t}\footnotesize{d\tilde{\Sigma}(p,t,t)\prod_{p,t>t'}d\text{Re}\tilde{\Sigma}(p,t,t')d\text{Im}\tilde{\Sigma}(p,t,t')}~e^{-\frac{N}{2\kappa^2}\sum_{p,t,t'}\tilde{A}(p)^{-1}\abs{\tilde{\Sigma}(p,t,t')}^2+\sum_{p,t,t'}\tilde{\Sigma}(p,t,t')\tilde{\xi}(p,t',t)}\ ,
\end{equation}
where $\tilde{\Sigma}(p)$ is a Hermitian matrix. By going back to $x$ representation, the right hand side may be written as an integral with respect to $\Sigma(x,t,t')$ fields:
\begin{equation}
e^{S_1}=\int\footnotesize{\prod_{x,t}d\Sigma(x,t,t)\prod_{x,t>t'}\text{Re}d\Sigma(x,t,t')d\text{Im}\Sigma(x,t,t')}~e^{-\frac{N}{2\kappa^2}\sum_{x,y,t,t'}\Sigma(x,t,t')A^{-1}(x,y)\Sigma(y,t',t)+\sum_{x,t,t'}\Sigma(x,t',t)\xi(x,t,t')}\ .
\end{equation}
Therefore, using (\ref{xi}) the action takes the form:
\begin{equation}\label{tilde_S}
\begin{aligned}
\tilde{S}&=\sum_{x,t,t',a}\bar{\psi}_a(x,t)\left[m\delta_{t,t'}+\hat{\gamma}_5(x)\hat{\partial}_t(t,t')+\Sigma(x,t',t)\right]\psi(x,t')_a\\
&-\frac{N}{2\kappa^2}\sum_{x,y,t,t'}\Sigma(x,t,t')A^{-1}(x,y)\Sigma(y,t',t)\ .
\end{aligned}
\end{equation}
Doing the Grassmann integral and using the identity $\det A=e^{\text{tr}\ln A}$ the final expression is:
\begin{equation}\label{sigma_action}
S_\Sigma=N\sum_{x,t}\left\{\ln\left[m+\hat{\gamma}_5(x)\hat\partial_t+\Sigma(x)\right]\right\}(t,t)-\frac{N}{2\kappa^2}\sum_{x,y,t,t'}\Sigma(x,t,t')(A^{-1})(x,y)\Sigma(y,t',t)\ .
\end{equation}
Note that $A$ is not invertible and positive definite and this poses a problem in the Gaussian integral representation. However, one may shift $A$ such as to be invertible and positive definite. The cost of doing so is the introduction of another Gaussian field. We have checked that, in this case, the saddle point solution is the same. Therefore, for simplicity we stay with a single Gaussian field and treat $A$ as being formally invertible and positive definite.

In this subsection we have shown that the dual theory is a matrix field theory. Since the action is proportional to $N$ one can solve this theory in the large $N$ approximation. Before doing so we define some useful observables in terms of matrix fields. We start with Green's functions of the theory.

\subsection{Green's functions}

Fermionic Green's function of the theory can be computed by adding the following Grassmann valued source terms in the action (\ref{I_action}):
\begin{equation}
{\cal I}(\chi,\bar\chi)={\cal I}+\sum_{x,t,a}[\bar\psi(x,t)_a\chi(x,t)_a+\bar\chi(x,t)_a\psi(x,t)_a]\ .
\end{equation}
For example, the two point function for $t\geq t'$, i.e.:
\begin{equation}\label{definition_greens_functions}
{\cal G}(x,y,t,t')_{ab}=\langle\psi(y,t')_a\bar\psi(x,t)_b\rangle_{\cal I}\ ,
\end{equation}
may be computed using the double derivative with respect to fermionic sources:
\begin{equation}
{\cal G}(x,y,t,t')_{ab}=\left.\frac{\partial^2}{\partial\bar\chi(y,t')_a\partial\chi(x,t)_b}\log Z(\chi,\bar{\chi})\right|_{\chi=\bar{\chi}=0}\ .
\end{equation}
Adding fermionic sources in the action $\tilde{S}$, equation (\ref{tilde_S}), and using the Grassmann integration rules, we get the updated $S_\Sigma$ action: 
\begin{equation}
S_\Sigma(\chi,\bar\chi)=S_\Sigma+\sum_{x,y,t,t',a,b}\bar\chi(x,t)_a\left[m+\hat{\gamma}_5(x)\hat\partial_t+\Sigma(x)\right]^{-1}_{(t,t')}\delta_{xy}\delta_{ab}\chi(y,t')_b\ .
\end{equation}
Therefore, the two point function is diagonal in color and $d$-dimensional space:
\begin{equation}\label{definition_greens_functions1}
{\cal G}(x,y,t,t')_{ab}=G(x,t,t')\delta_{xy}\delta_{ab}\ ,
\end{equation}
where:
\begin{equation}
G(x,t,t')=\left[m+\hat{\gamma}_5(x)\hat\partial_t+\Sigma(x)\right]^{-1}_{(t,t')}\ .
\end{equation}
Applying the same procedure for the four-point function we find:
\begin{equation}\label{four_point_function}
\langle\psi(x,t')_a\bar\psi(x,t)_a\psi(y,t)_b\bar\psi(y,t')_b\rangle_{\cal I}=G(x,t,t')G(y,t',t)\ .
\end{equation}
In the next subsection we turn to the Polyakov loop.

\subsection{Polyakov loop}

A key observable in lattice gauge theory is the Polyakov loop:
\begin{equation}
{\cal P}(\vec{x})=\left\langle\frac{1}{N}\text{~Tr}_{SU(N)}\prod_{x_4=1}^{N_4}U_4(\vec{x},x_4)\right\rangle\ ,
\end{equation}
where the expectation value is evaluated with respect to the effective theory (\ref{effective_action_h2}). In the following we express the expectation value with respect to degrees of freedom of the dual theory. Adding the fermionic source term in the action (\ref{I_action}):
\begin{equation}
{\cal I}(a,\bar{a},b,\bar{b})={\cal I}-\kappa\sum_{x,\mu}\eta_\mu(x)\left[\bar{a}(x)U_{\mu}(x)a(x+\hat{\mu})-\bar{b}(x+\hat{\mu})U_{\mu}(x)^*b(x)\right]\ ,
\end{equation}
then, the insertion of gauge fields in the path integral may be achieved using the double derivative of the partition function with respect to source fields:
\begin{equation}
\left.\frac{\partial^2}{\partial a(x+\hat{\mu})_j\partial \bar{a}(x)_i}\log Z(a,\bar{a},b,\bar{b})\right|_{a=\bar{a}=b=\bar{b}=0}=\left\langle -\kappa\eta_\mu(x)U_{\mu}(x)_{ij}\right\rangle_{\cal I}\ ,
\end{equation}
where $i,j$ are SU(N) matrix indices and the expectation is evaluated with respect to the action of eq. (\ref{I_action}). After integration of gauge fields and using the same approximation that led us to the action (\ref{psi_action}) one gets the same action supplemented with source terms:
\begin{equation}
\begin{aligned}
&S(a,\bar{a},b,\bar{b})=S+\frac{\kappa^2}{N}\sum_{x,\mu,t}\left[\bar{\psi}(x+\hat{\mu},t)a(x+\hat{\mu})\bar{a}(x)\psi(x,t)+\bar{\psi}(x,t)b(x)\bar{b}(x+\hat{\mu})\psi(x+\hat{\mu},t)\right]\\
&+\frac{\kappa^2}{N}\sum_{x,\mu}\bar{a}(x)b(x)\bar{b}(x+\hat{\mu})a(x+\hat{\mu})\ .
\end{aligned}
\end{equation}
The effect of the double derivative in this theory is now the insertion of the following bilinear combination of fermionic fields:
\begin{equation}
\left.\frac{\partial^2}{\partial a(x+\hat{\mu})_j\partial \bar{a}(x)_i}\log Z(a,\bar{a},b,\bar{b})\right|_{a=\bar{a}=b=\bar{b}=0}=\left\langle -\frac{\kappa^2}{N}\sum_t{\psi}(x,t)_i\bar{\psi}(x+\hat{\mu},t)_j\right\rangle_S\ ,
\end{equation}
where now the expectation value is evaluated in the theory defined by the action of eq. (\ref{psi_action}). Therefore, the Polyakov loop may be written in the form:
\begin{equation}
{\cal P}(\vec{x})=\sum_{t_1,\ldots,t_{N_4}}\left\langle\prod_{x_4=1}^{N_4}\sum_{j_{x_4}=1}^N\bar{\psi}(\vec{x},x_4,t_{x_4-1})_{j_{x_4}}\psi(\vec{x},x_4,t_{x_4})_{j_{x_4}}\right\rangle_S\ ,
\end{equation}
modulo an irrelevant constant factor and where $t_o=t_{N_4}$. By means of two point functions this expression may be written in the form:
\begin{equation}\label{polyakov_loop}
{\cal P}(\vec{x})=\sum_{t_1,\ldots,t_{N_4}}\left\langle\prod_{x_4=1}^{N_4}G(\vec{x},x_4,t_{x_4-1},t_{x_4})\right\rangle_{S_\Sigma}\ ,
\end{equation}
modulo a trivial factor that comes from color summations. In the next subsection, we define one more observable, the fermion-antifermion condensate.

\subsection{Fermion-antifermion condensate}

An important observable in a theory of fermions is the fermion-antifermion condensate:
\begin{equation}
\zeta=\lim_{V\rightarrow\infty}\frac{1}{V}\sum_{x,t}\left\langle\sum_{j=1}^N\bar{\psi}(x,t)_j\psi(x,t)_j\right\rangle_S\ .
\end{equation}
From the discussion in the previous subsection we conclude that:
\begin{equation}\label{qqbar}
\zeta=\lim_{V\rightarrow\infty}\frac{N}{V}\sum_{x,t}\left\langle G(x,t,t) \right\rangle_{S_\Sigma}\ .
\end{equation}
In this study we restrict ourselves in this limited set of observables. In the following, we turn to the solution of the theory in the large N limit.

\section{The large $N$ solution}
\label{solution1}

Since the action, equation (\ref{sigma_action}), is proportional to $N$, one may employ the saddle point solution of the theory. The stationary field should satisfy the necessary first order conditions, which in our case is the system of equations:
\begin{equation}
\frac{\partial S_\Sigma}{\partial\Sigma(x,t',t)}=0\ ,~~~~~~~~t,t'=1,2\ldots,N_t
\end{equation}
for any lattice site $x$ of the $d$-dimensional lattice. The derivative of the first term is taken by expanding the matrix logarithm as a power series on $\Sigma$. This way, we obtain the system of equations:
\begin{equation}
G(x,t',t)=\frac{1}{\kappa^2}\sum_y(A^{-1})(x,y)\Sigma(y,t',t)\ .
\end{equation}
These can be inverted to give:
\begin{equation}\label{sigma_solution}
\Sigma(x,t',t)=\kappa^2\sum_{\mu}[G(x+\hat{\mu},t',t)+G(x-\hat{\mu},t',t)]\ .
\end{equation}
Therefore, we get a coupled system of quadratic equations for Green's functions:
\begin{equation}
G^{-1}(x,t',t)=\delta_{t,t'}+\hat{\gamma}_5(x)\hat{\partial}_t(t,t')+\kappa^2\sum_{\mu}[G(x+\hat{\mu},t',t)+G(x-\hat{\mu},t',t)]\ .
\end{equation}
We are interested in the vanishing $\kappa$ solution of the theory. In this limit, the second term of the right hand side is small. Given the translation invariance of the time derivative the small $\kappa$ limit of the solution is also translation invariant, i.e. $G(x,t',t)$ is a function of time separation $t'-t$. Therefore, equations may be written in terms of Fourier transformed Green's functions $\hat{G}(x,\omega)$:
\begin{equation}\label{G_eq}
\frac{1}{\hat{G}(x,\omega)}=m+\hat{\gamma}_5(x)i\sin\omega+\kappa^2\sum_{\mu}[\hat{G}(x+\hat{\mu},\omega)+\hat{G}(x-\hat{\mu},\omega)]
\end{equation}
for each frequency $\omega$. We solve the system using the solution Ansatz:
\begin{equation}\label{ansatz}
\hat{G}(x,\omega)=e^{-i\arg\left[m+\hat{\gamma}_5(x)i\sin\omega\right]}\left[\tilde{G}_o(\omega)+\tilde{G}(x,\omega)\right]\ ,
\end{equation}
where $\tilde{G}(x,\omega)$ is a small fluctuation around $x$-independent solution $\tilde{G}_o(\omega)$. Expanding the left hand side of the system (\ref{G_eq}) up to the first order in $\tilde{G}(x,\omega)$ and using the fact that (\ref{G_eq}) relates the solution on even and odd sites, we get a scalar quadratic equation:
\begin{equation}\label{G_o_eq}
\frac{1}{\tilde{G}_o(\omega)}=\abs{m+i\sin\omega}+2d\kappa^2\tilde{G}_o(\omega)\ ,
\end{equation}
as well as the first order constraint:
\begin{equation}\label{first_order_constraint}
\tilde{G}(x,\omega)+\kappa^2\tilde{G}_o(\omega)^2\sum_{\mu}[\tilde{G}(x+\hat{\mu},\omega)+\tilde{G}(x-\hat{\mu},\omega)]=0\ .
\end{equation}
Substituting the Fourier transformed equation (\ref{sigma_solution}) in the action (\ref{sigma_action}) and writing it in terms of $\hat{G}(x,\omega)$ we get:
\begin{equation}
S_{\hat{G}}/N=-\sum_{x,\omega}\ln\hat{G}(x,\omega)-\frac{\kappa^2}{2}\sum_{x,y,\omega}\hat{G}(x,\omega)A(x,y)\hat{G}(y,\omega)\ .
\end{equation}
Expanding the logarithm up to the second order in $\tilde{G}_o^{-1}\tilde{G}(x,\omega)$ and substituting the constraint (\ref{first_order_constraint}) we get the effective action in the large $N$ approximation:
\begin{equation}\label{eff_action_large_N}
\begin{aligned}
S_{\tilde{G}}/N&=-V\sum_{\omega}\left[\ln\tilde{G}_o(\omega)+d\kappa^2 \tilde{G}_o(\omega)^2\right]\\
&+\frac{\kappa^2}{2}\sum_{x,\omega}\left(\frac{1}{\kappa^2\tilde{G}_o(\omega)^2}-2d\right)\tilde{G}(x,\omega)^2+\frac{\kappa^2}{2}\sum_{x,\mu,\omega}\left[\tilde{G}(x+\hat{\mu},\omega)-\tilde{G}(x,\omega)\right]^2\\
&+O\left[\sum_\omega\tilde{G}_o(\omega)^{-3}\tilde{G}(x,\omega)^3\right]\ ,
\end{aligned}
\end{equation}
where translation symmetry of $\tilde{G}(x,\omega)$ on the $d$ dimensional lattice has been used, i.e. $\sum_{x}\tilde{G}(x+\hat{\mu},\omega)^2=\sum_{x}\tilde{G}(x,\omega)^2$. Neglecting cubic order corrections this action describes a free theory of $N_t$ bosons with masses given by:
\begin{equation}\label{mass_equation}
M(\omega)^2=\frac{1}{\kappa^2\tilde{G}_o(\omega)^2}-2d\ .
\end{equation}
The symmetry of the action can be made explicit by writing it in the form of a non-linear sigma model:
\begin{equation}\label{nonlinear_sigma_model}
S_{\tilde{G}}/N\propto-\sum_{x,\omega}M(\omega)^2e^{i\tilde{G}(x,\omega)}-\sum_{x,\mu,\omega}e^{i\tilde{G}(x+\hat{\mu},\omega)}e^{-i\tilde{G}(x,\omega)}+O\left[\tilde{G}_o(\omega)^{-3}\tilde{G}(x,\omega)^3\right]+h.c.\ .
\end{equation}
If we define $\tilde{G}(x)$ matrices with matrix elements $\tilde{G}(x,\omega)\delta_{\omega,\omega'}$, then the global unitary transformations:
\begin{equation}
e^{i\tilde{G}(x)}\rightarrow Ue^{i\tilde{G}(x)}U^*\ ,~~~~~~~~U\in U(N_t)\ ,
\end{equation}
with $U(N_t)$ denoting the unitary group, leave the spectrum of the theory invariant. In the following subsection we compute the free energy and the mass gap of the theory.

\subsection{Free energy and mass gap}

We compute the solution $\tilde{G}_o(\omega)$ of the quadratic equation (\ref{G_o_eq}). Denoting:
\begin{equation}
\mu(\omega)=\abs{m+i\sin\omega}\ ,
\end{equation}
we get:
\begin{equation}\label{tilde_G_o}
\tilde{G}_o(\omega)=\frac{-\mu(\omega)+\sqrt{\mu(\omega)^2+8d\kappa^2}}{4d\kappa^2}\ .
\end{equation}
Using equation (\ref{eff_action_large_N}) the free energy reads:
\begin{equation}
{\cal F}(\kappa)=NV\sum_{\omega}\left[\ln\tilde{G}_o(\omega)+d\kappa^2 \tilde{G}_o(\omega)^2\right]\ ,
\end{equation}
whereas substituting $\tilde{G}_o(\omega)$ in the equation (\ref{mass_equation}) the mass spectrum of the theory is given by:
\begin{equation}\label{mass_spectrum}
M(\omega)^2=\frac{\mu(\omega)^2+\mu(\omega)\sqrt{\mu(\omega)^2+8d\kappa^2}}{2\kappa^2}\ ,
\end{equation}
The spectrum is bounded from below by the mass gap:
\begin{equation}\label{mass_gap}
M_o^2=\frac{m^2+m\sqrt{m^2+8d\kappa^2}}{2\kappa^2}\ .
\end{equation}
In the case of massive fermions and vanishing $\kappa$ the mass gap $M_o$ diverges as $m/\kappa$. In the following subsection we study Green's functions of the theory.

\subsection{Two point function}

Using the solution Ansatz (\ref{ansatz}) the leading order approximation of the two point function is:
\begin{equation}
\tilde{G}_{\pm}(\omega)=e^{-i\arg\left(m\pm i\sin\omega\right)}\tilde{G}_o(\omega)\ ,
\end{equation}
where we have used $\hat{\gamma_5}(x)=\pm 1$ on even/odd sites. Substituting $\tilde{G}_o(\omega)$ from equation (\ref{tilde_G_o}) and expanding the square root in $\kappa$ for massive fermions we get:
\begin{equation}\label{G_plus_minus}
\tilde{G}_\pm(\omega)=\frac{1}{m\pm i\sin\omega}+O(\kappa^2)\ .
\end{equation}
In the following we concentrate on even sites two-point function, i.e. $\hat{\gamma_5}(x)=1$ and compute the first term of the right hand side. The odd sites expression is then the negative even sites result with the formal substitution $m\to-m$. The time domain two-point function may be computed using the inverse discrete Fourier transform:
\begin{equation}\label{inv_fourier}
X_+(t)=\frac{1}{N_t}\sum_\omega\frac{e^{it\omega}}{m+i\sin\omega}
\end{equation}
plus $O(\kappa^2)$ terms. The leading term $X_+(t)$ is thus the propagator of free fermions in 0+1 dimensions. The latter may be obtained by solving the linear system of equations:
\begin{equation}
\left(m+\hat\partial_t\right)X_+(t)=\delta_{t,0}\ ,~~~~~~~~t=0,1,\ldots,N_t-1
\end{equation}
with antiperiodic boundary conditions. Both ways we find:\footnote{The right hand side of (\ref{inv_fourier}) may be computed in two steps: first, one computes the $N_t\to\infty$ expression, i.e. $X_+^{(\infty)}(t)=\int_{-\pi}^{\pi}\frac{d\omega}{2\pi}\frac{e^{it\omega}}{m+i\sin\omega}=\frac{e^{-E\abs{t}}}{\cosh E}[\Theta(t)+(-1)^t\Theta(-t)]$, where $\Theta(t)$ is the Heaviside function. Then, the finite $N_t$ result is obtanied by evaluating the infinite sum $X_+(t)=\sum_{m=-\infty}^{+\infty}(-1)^{\abs{m}}X_+^{(\infty)}(t+mN_t)$.}
\begin{equation}\label{two_point_function_massive}
X_+(t)=\frac{1}{C}\left\{
\begin{matrix}
\sinh E(N_t/2-t) & ~~~~t\in\text{even}\ ,\\
\cosh E(N_t/2-t) & ~~~~t\in\text{odd}
\end{matrix}
\right.
\end{equation}
with $\sinh E=m$, $C=\cosh E\cosh(EN_t/2)$ and $N_t$ even. In case $N_t$ is odd the above result should be replaced by the expression:
\begin{equation}\label{two_point_function_massive1}
X_+(t)=\frac{1}{S}\left\{
\begin{matrix}
\cosh E(N_t-t)+\sinh Et & ~~~~t\in\text{even}\ ,\\
\sinh E(N_t-t)-\cosh Et & ~~~~t\in\text{odd}
\end{matrix}
\right.
\end{equation}
with $S=\cosh E\sinh(EN_t)$. The different behavior of the two point function at even and odd times is a manifestation of fermion doubling on the lattice, i.e. the presence of two poles of $1/(m+i\sin\omega)$ at $iE$ and $\pi-iE$ corresponding to the same energy $E$.

\subsection{Polyakov loop}

We begin by expressing the time domain trace of (\ref{polyakov_loop}) in the frequency domain:
\begin{equation}
{\cal P}(\vec{x})=\sum_\omega\left\langle\prod_{x_4=1}^{N_4}G(\vec{x},x_4,\omega)\right\rangle_{S_\Sigma}\ .
\end{equation}
In the large $N$ approximation the right hand side factorizes and we find:
\begin{equation}
{\cal P}(\vec{x})=
\left\{
\begin{matrix}
\sum_\omega \tilde{G}_o(\omega)^{N_4} & ~~~~N_4\in\text{even}\ ,\\
\sum_\omega e^{-i\arg\left(m\pm i\sin\omega\right)}\tilde{G}_o(\omega)^{N_4} & ~~~~N_4\in\text{odd}~~~~
\end{matrix}
\right.
\end{equation}
since for $N_4$ even phase factors cancel due to equal number of even and odd sites in the product. If $N_4$ is odd then only one phase factor remains. Due to the $\pi$-periodicity of $\tilde{G}_o(\omega)$ the Polyakov loop may be written as a sum over positive frequency terms:\footnote{This is true in case there is an even number of frequencies, otherwise one should add an extra contribution coming from $\omega=\pi$, which does not change the result.}
\begin{equation}\label{polyakov_large_N}
{\cal P}(\vec{x})=
\left\{
\begin{matrix}
2\sum_{\omega>0}\tilde{G}_o(\omega)^{N_4} & ~~~~N_4\in\text{even}\ ,\\
2\sum_{\omega>0} \frac{1}{\sqrt{m^2+\sin^2\omega}}\tilde{G}_o(\omega)^{N_4} & ~~~~N_4\in\text{odd}\ .
\end{matrix}
\right.
\end{equation}
We evaluate the sum in the large $N_4$ limit, in which case the largest term dominates. Note that $\tilde{G}_o(\omega)$ is maximum at $\omega=\pi/N_t$. In the large $N_t$ limit $\pi/N_t$ is close to zero and we evaluate:
\begin{equation}\label{polyakov_loop_result_ym}
{\cal P}(\vec{x})\simeq 2N_t\tilde{G}_o(0)^{N_4}\ .
\end{equation}
This way, the free energy $F_o$ of the static charge is:
\begin{equation}\label{F_o}
aF_o=-\ln\tilde{G}_o(0)\ ,
\end{equation}
where we have restored the lattice spacing $a$. At leading order in $\kappa$ and $m=1$ we have $\tilde{G}_o(0)=1-2d\kappa^2+O(\kappa^4)$ and therefore:
\begin{equation}
F_o=\frac{1}{a}2d\kappa^2+O(\kappa^4)\ ,
\end{equation}
which vanishes in the limit $\kappa\rightarrow 0$. The result does not change if we use $\tilde{G}(\pi/N_t)$ instead of $\tilde{G}(0)$ as well as if we add the next largest term in the sum over frequencies in (\ref{polyakov_large_N}).

Using $F_o$ we can compute the renormalization group the beta function:
\begin{equation}
\beta(\kappa)=-a\frac{d\kappa}{da}=\frac{a~(\partial F_o/\partial a)}{(\partial F_o/\partial\kappa)}=-\frac{\kappa}{2}+O(\kappa^3)\ .
\end{equation}
It is negative and vanishes linearly with the coupling constant. The theory is thus asymptotically free. However, its ultraviolet behavior is different from the standard Yang-Mills theory. The same conclusion may be drawn if we had computed Wilson loops. In this case we would find a perimeter law.

\subsection{Fermion-antifermion condensate}

Using its definition, equation (\ref{qqbar}), and substituting the two point function in the leading order approximation of the large $N$ solution, the condensate is:
\begin{equation}\label{qqbar_1}
\zeta=N\sum_\omega e^{-i\arg\left(m+i\sin\omega\right)}\tilde{G}_o(\omega)=2N\sum_{\omega>0}\frac{m}{\sqrt{m^2+\sin^2\omega}}\tilde{G}_o(\omega)\ ,
\end{equation}
where we have used again the $\pi$-periodicity of $\tilde{G}_o(\omega)$. Expanding the right hand side in $\kappa$ and taking $m=1$ we have:
\begin{equation}
\zeta=N\sum_\omega\frac{1}{1+\sin^2\omega}+O(\kappa^2)\ .
\end{equation}
For large $N_t$ the first term is a lattice sum equal to $N_t/\sqrt{2}$ \footnote{In this limit one may compute the integral $N_t\int_{-\pi}^{\pi}d\omega/(2\pi)/(1+\sin^2\omega)$.} and therefore:
\begin{equation}
\zeta=NN_t\frac{1}{\sqrt{2}}+O(\kappa^2)\ .
\end{equation}
This way, in the continuum limit, the theory is characterized by a non-zero value of the fermion-antifermion condensate.

\subsection{Synthesis}

The solution of the theory in the large $N$ approximation shares distinctive properties with the standard Yang-Mills theory like asymptotic freedom and color confinement. Note however, that the beta function of our effective Yang-Mills theory is different from that of the standard Yang-Mills theory. In the next section we probe the theory in the massless limit.

\section{Asymptotic safe QCD}
\label{solution2}

We have seen that the effective Yang-Mills theory is local and asymptotically free. In this section we compute the solution in the case of massless fermions, or almost massless fermions with mass $m$. In subsection \ref{massless_fermions} we learned that the effective theory of our model is QCD with a large number of flavors. The large $N$ solution is found following the same steps as in the previous section. The mass spectrum of the theory is given by equation (\ref{mass_spectrum}) with:
\begin{equation}
\mu(\omega)^2=m^2+\sin^2\omega\ .
\end{equation}
However, the lightest mass:
\begin{equation}\label{lightest_mass}
M_o^2=\frac{m^2+m\sqrt{m^2+8d\kappa^2}}{2\kappa^2}
\end{equation}
vanishes for exactly massless fermions. Therefore, the theory is gapless. Note however, that the number of light fermions is large, as can be seen  from equation (\ref{fermion_masses}). If there are $N_f$ such modes, then the effective action of the theory, equation (\ref{nonlinear_sigma_model}), may be split in two pieces corresponding to light and heavy modes:
\begin{equation}
\begin{aligned}
S_{\tilde{G}}/N\propto&-\sum_{x,\omega\in\Omega_l}M(\omega)^2e^{i\tilde{G}(x,\omega)}-\sum_{x,\mu,\omega\in\Omega_l}e^{i\tilde{G}(x+\hat{\mu},\omega)}e^{-i\tilde{G}(x,\omega)}\\
&-\sum_{x,\omega\in\Omega_h}M(\omega)^2e^{i\tilde{G}(x,\omega)}-\sum_{x,\mu,\omega\in\Omega_h}e^{i\tilde{G}(x+\hat{\mu},\omega)}e^{-i\tilde{G}(x,\omega)}+O\left[\tilde{G}_o^{-3}\tilde{G}(x)^3\right]+h.c.
\end{aligned}
\end{equation}
with $\Omega_l$ and $\Omega_h$ being the set of light and heavy mode frequencies. If we define light mode matrices $\tilde{G}_l(x)$ with matrix elements:
\begin{equation}
\tilde{G}_l(x,\omega,\omega')=\tilde{G}(x,\omega)\delta_{\omega,\omega'}\ ,~~~~~~~~\omega,\omega'\in\Omega_l
\end{equation}
and take light modes to be massless, the action is symmetric with respect to global $U(N_f)_L\times U(N_f)_R$ chiral transformations of light modes:
\begin{equation}
e^{i\tilde{G}_l(x)}\rightarrow Ue^{i\tilde{G}_l(x)}V^*\ ,~~~~~~~~U,V\in U(N_f)\ .
\end{equation}
As shown below, the fermion-antifermion condensate of the full theory is nonzero due to light modes alone. Therefore, the chiral symmetry of the light modes is spontaneously broken to $U(N_f)$. Taking the limit of vanishing $m$ in equation (\ref{lightest_mass}) the mass of the $N_f^2$ Goldstone modes is:
\begin{equation}\label{boson_goldstone_mass}
M_o^2\simeq\frac{m}{\kappa}\sqrt{2d}\ ,
\end{equation}
a result which is expected from the chiral perturbation theory \cite{gasser_leutwyler}. Therefore, the solution shows that the low lying spectrum behaves as in QCD. Note however that $N_f$ is large in our case.

\subsection{Two point function}

We compute the two point function using again the solution Ansatz (\ref{ansatz}) in the leading order approximation. The massless two point function is then:
\begin{equation}
\tilde{G}_{\pm}(\omega)=\pm i~\text{sgn}(\omega)\tilde{G}_o(\omega)\ ,
\end{equation}
where the plus/minus subscript corresponds to even/odd sites of the $d$ dimensional lattice. Substituting $\tilde{G}_o(\omega)$ from equation (\ref{tilde_G_o}) we get:
\begin{equation}
\tilde{G}_{\pm}(\omega)=\pm\frac{-i\sin\omega+i~\text{sgn}(\omega)\sqrt{\sin^2\omega+8d\kappa^2}}{4d\kappa^2}\ .
\end{equation}
In continuous time, this is the same as the two point function of the $q=2$ SYK model, i.e.:
\begin{equation}
\tilde{G}_+^{cont}(\omega)=\frac{-i\omega+i~\text{sgn}(\omega)\sqrt{\omega^2+4J^2}}{2J^2}\ ,
\end{equation}
where $J$ is the coupling constant of the SYK model, with the identification $J=\sqrt{2d}\kappa$. The time domain two point function is given by Maldacena and Stanford \cite{maldacena_stanford}:
\begin{equation}
X_+^{cont}(t)=\text{sgn}(t)\int_0^\pi\frac{d\theta}{\pi}\cos^2\theta~e^{-2\sqrt{2d}\kappa\abs{t}\sin\theta}\ ,
\end{equation}
which for large time separations is:
\begin{equation}\label{two_point_function_massless}
X_+^{cont}(t)=\frac{1}{\pi\sqrt{2d}\kappa t}-\frac{1}{4\pi(\sqrt{2d}\kappa t)^3}+O(\kappa^{-5}t^{-5})\ .
\end{equation}
This is a power law decay as opposed to exponential decay in the case of massive fermions.

\subsection{Asymptotic safety}

Following the same steps as in the case of massive fermions we compute the Polyakov loop, equation (\ref{polyakov_large_N}), in the large $N_4$ limit. The free energy of the static charge is thus evaluated at zero frequency using the same formula (\ref{F_o}). The result in the massless case is:
\begin{equation}
aF_o=\ln\sqrt{2d}\kappa\ .
\end{equation}
Using $F_o$, the beta function of the theory is:
$$
\beta(\kappa)=-\kappa\ln\left(\sqrt{2d}\kappa\right)\ .
$$
It is zero at $\kappa=0$ and $\kappa_c=1/\sqrt{2d}$, positive for $\kappa\in(0,\kappa_c)$ and negative for $\kappa>\kappa_c$. Hence, the theory has an ultraviolet fixed point at $\kappa_c$. Solving the renormalization group equation:
$$
-a\frac{d\kappa}{da}=-\kappa\ln\left(\sqrt{2d}\kappa\right)\ ,~~~~~~~~\Rightarrow~~~~~~~~a\tilde{m}=\ln\left(\sqrt{2d}\kappa\right)\ ,
$$
with $\tilde{m}$ being an integration constant, the correlation length of the theory is defined by:
$$
\xi=\frac{1}{a\tilde{m}}\ .
$$
At the critical point, it diverges according to the law:
$$
\xi\propto\left|1-\frac{\kappa}{\kappa_c}\right|^{-1}\ ,
$$
i.e. the theory shares the same critical exponent with two-dimensional Ising model. The theory has a continuum limit at a critical value of the coupling constant. Note that in the limit $d\rightarrow\infty$ the theory becomes asymptotically free.

There are some consequences of the critical theory. For exmaple, at critical $\kappa$ the mass of the Goldstone boson, equation (\ref{boson_goldstone_mass}) becomes:
\begin{equation}
M_o^2\simeq 2dm\ ,
\end{equation}
which again vanishes as expected from chiral perturbation theory. Another consequence is that the length of the extra dimension is finite in the continuum limit. However, due to relation (\ref{N_t_kappa}), i.e. $N_t=4\kappa^{-5}$ the critical value of $N_t$ is $N_t^{(c)}=4(2d)^{5/2}$. In the interesting case $d=4$ the value $N_t^{(c)}\approx 724$, which is indeed large. Even at $d=2$ we have $N_t^{(c)}=128$. Therefore, both theories considered in this paper, for massive and massless fermions, satisfy the large $N_t$ assumption which has been used by us from the beginning.

\subsection{Fermion-antifermion condensate}

Splitting the sum in the equation (\ref{qqbar_1}) in to light and heavy frequency contributions we find:
\begin{equation}
\zeta(m)\approx 2NN_f\tilde{G}_o(0)+O(m)\ .
\end{equation}
Therefore, in the massless limit we get:
\begin{equation}
\zeta(0)\approx 2NN_f\frac{1}{\sqrt{2d}\kappa}\ .
\end{equation}
This way, in the continuum limit, the theory has a non-zero chiral condensate. At critical $\kappa$ its value is independent of the coupling constant and the number of dimensions.

\section{Relation to $q=2$ SYK model}
\label{syk2}

Recently, there is a great deal of work on discovering solvable examples of the AdS/CFT correspondence. Such an example is the SYK model. Its effective action on a time lattice has the form:\footnote{See for example equation A.9 of Gross and Rosenhaus paper \cite{gross_rosenhaus}.}
\begin{equation}\label{syk_q}
-S_{SYK,q}=\frac{N}{2}\sum_t\left[\ln\left(\hat\partial_t-\Sigma\right)\right](t,t)+\frac{1}{2}\left\{\frac{J^2N}{q}\sum_{t,t'}G(t,t')^q-N\sum_{t,t'}G(t,t')\Sigma(t,t')\right\}\ ,
\end{equation}
where $G(t,t')$ and $\Sigma(t,t')$ are bilocal fields defined on a time lattice, $N$ is a large positive integer and $J$ is the coupling constant of the theory. In this subsection we relate our model, i.e. equation (\ref{sigma_action}):
\begin{equation}
S_\Sigma=N\sum_{x,t}\left\{\ln\left[m+\hat{\gamma}_5(x)\hat\partial_t+\Sigma(x)\right]\right\}(t,t)-\frac{N}{2\kappa^2}\sum_{x,y,t,t'}\Sigma(x,t,t')(A^{-1})(x,y)\Sigma(y,t',t)
\end{equation}
to the $q=2$ SYK model. The relation is established in the massless case under the following conditions:
\begin{enumerate}[i)]
\item the large $N$ limit;
\item the coupling constant relationship $J^2=2d\kappa^2$.
\end{enumerate}
The saddle point solution $\tilde{G}_o(\omega)$, which is $x$ independent, suggests that we may approximate the matrix $A$ by a constant matrix, i.e. $A=2d$. This way, the action (\ref{sigma_action}) decouples completely in $x$-space:
\begin{equation}
S_\Sigma=N\sum_{x,t}\left\{\ln\left[m+\hat{\gamma}_5(x)\hat{\partial}_t+\Sigma(x)\right]-\frac{1}{4d\kappa^2}\Sigma(x)^2\right\}(t,t)\ .
\end{equation}
This is an ideal gas of pairs of one-matrix theories:
\begin{equation}
S_{\Sigma,\pm}=N\sum_t\left[\ln\left(m\pm\hat{\partial}_t+\Sigma\right)-\frac{1}{4d\kappa^2}\Sigma^2\right](t,t)
\end{equation}
corresponding to even/odd sites of the $d$ dimensional lattice. Picking even sites only and setting $m=0$ we end up with model:
\begin{equation}\label{sigma_1_action}
S_{\Sigma,1}=N\sum_t\left[\ln\left(\hat{\partial}_t+\Sigma\right)-\frac{1}{4d\kappa^2}\Sigma^2\right](t,t)\ .
\end{equation}
One-matrix models have been studied in the past as a non-perturbative formulation of the two-dimensional gravity, see for example \cite{ginsparg}. In general, such models do not lead to black hole formation \cite{karczmarek_et_al}. As it was shown in the previous section, the theory with massless fermions  shares the same two-point function with the action of $q=2$ SYK model. This is not a coincidence. The action (\ref{sigma_1_action}) may be written in terms of an additional Hermitian matrix $G(t,t')$:
\begin{equation}
S_{\Sigma,G,1}=N\sum_t\left[\ln\left(\hat\partial_t+\Sigma\right)\right](t,t)-\frac{2d\kappa^2N}{2}\sum_{t,t'}G(t,t')G(t',t)+iN\sum_{t,t'}G(t,t')\Sigma(t,t')\ ,
\end{equation}
which can be shown by integrating $e^{S_{\Sigma,G,1}}$ with respect to individual matrix elements $G(t,t')$ using Gaussian integrals (\ref{gaussian1}) and (\ref{gaussian2}). Rescaling the matrix $G\rightarrow-iG$ as well as the matrix $\Sigma\rightarrow-\Sigma$ we get twice the action of the $q=2$ SYK model on a time lattice:
\begin{equation}
S_{\Sigma,G,1}=N\sum_t\left[\ln\left(\hat\partial_t-\Sigma\right)\right](t,t)+\frac{2d\kappa^2N}{2}\sum_{t,t'}G(t,t')G(t',t)-N\sum_{t,t'}G(t,t')\Sigma(t,t')\ ,
\end{equation}
where $2d\kappa^2=J^2$ is identified with the square the coupling constant of the SYK model. Therefore, the large $N$ asymptotic safe QCD may be described as an ideal gas of $q=2$ SYK models on each site of a $d$ dimensional lattice. Note that since $d$ and $\kappa$ are related in continuum limit such that $\kappa_c=1/\sqrt{2d}$ we have $J^2_c=1$. Therefore, the relationship (\ref{N_t_kappa}), i.e. $N_t=4\kappa^{-5}$, has no influence on the magnitude of $J$. It merely tells that at critical $\kappa$ the length of the extra dimension $N_t^{(c)}=4(2d)^{5/2}$ is large. Therefore, the continuum limit of the asymptotic safe QCD corresponds to the low temperature limit of the $q=2$ SYK model with $J^2=1$.

Note that for $q=2$ the SYK model is not chaotic, whereas for $q=4$ it saturates the chaos bound. Since we used the leading order approximation of the $F$ function (see equation (\ref{F_function})), the effect of order $q=4$ terms or higher is expected to be present in the effective action. The extent to which these terms alter the chaotic behavior of our theory remains unclear without a proper calculation. However, the $q=2$ SYK model originates from a quadratic Hamiltonian with disorder couplings. Magan has shown that a generic model of quadratic fermions with random couplings satisfies the Eigenstate Thermalization Hypothesis \cite{magan}.

\section{Summary and discussion}

In this paper, we have formulated and studied a lattice theory of fermions beyond the Standard Model. Integration of fermions yields a lattice gauge theory which is expressed in terms of Wilson loops of growing sizes. This premise is interesting alone since the effective theory is a local Yang-Mills theory in the limit of vanishing coupling constant. On the other hand the fermion theory may be integrated with the help of known one-link integrals. The remaining effective theory of fermions is then bosonized with the help of Hermitian matrices. The dual theory obtained this way is solved in the large $N$ limit. The solvability of the theory is a distinctive property of the model. The solution shows that the model shares qualitative properties of strong interactions like asymptotic freedom, color confinement and a spectral gap. However, the renormalization group beta function of the theory vanishes linearly with the coupling constant as opposed to the cubic law of the standard Yang-Mills theory \cite{gross_wilczek,politzer}. Nonetheless, the main result of the paper is that a local Yang-Mills theory exists which is non-perturbatively solvable in contrast to the present status of an unknown similar solution to the standard Yang-Mills theory in four dimensions.

We have studied also the model with massless fermions. In this case the effective theory of Yang-Mills fields is QCD with a large number of light flavors. Its solution is ultraviolet complete at a non-zero critical coupling constant where the theory is scale invariant. The light modes of the theory are shown to be chirally symmetric, a symmetry which is spontaneously broken. The theory shares the same two point function with the $q=2$ SYK model in the leading order of large $N$ approximation and the leading order approximation of our bosonization approach.

In the appendix \ref{gaussian_disorder} we have shown that one may use complex $N\times N$ matrices as disorder couplings instead of SU(N) couplings. Within the approximation made for the bosonization of fermions in case of the SU(N) disorder and the scaling relation assumed between $N$ and $N_t$ (see appendix \ref{gaussian_disorder}) we show that SU(N) and Gaussian disorder give closely related effective theories.

The results of this paper show that the model has a rich structure, which we intend to study further in the future. The next step is to treat the bosonization of fermions exactly. In order to probe the theory further we would like to compute more physical quantities such as the low lying meson spectrum. The computation of fermion-antifermion potential would reveal the nature of the interactions of the theory. We would like to study further the connection to the SYK model with the intention to find whether the theory has a gravity dual. Finally, we would like also to simulate the theory on the computer.

\subsection*{Acknowledgements}

The author would like to thank Michael Creutz, Philippe de Forcrand and Martin L\"uscher for comments and correspondence on the first draft of the paper. The expanded version of this work has benefited from the reports of the paper reviewer.

\pagebreak

\appendix

\section{Gaussian disorder}
\label{gaussian_disorder}

The SU(N) disordered fields are a special case of more general model of lattice fermions coupled to general $N\times N$ complex matrices $\Phi_\mu(x)$ attached at each directed link $(x,x+\hat{\mu})$ on the lattice. Adopting the same notations as in subsection (\ref{hamilton_operator}) the Hamiltonian operator of the theory is:
\begin{equation}\label{phi_hamiltonian}
\begin{aligned}
H&=\sum_{x,a}\Psi(x)^*_a\hat{\gamma}_5(x)\Psi(x)_a\\
&+\kappa\sum_{x,a,b,\mu}\hat{\gamma}_5(x)\eta_\mu(x)\left[\Psi(x)_a^*\Phi_{\mu}(x)_{ab}\Psi(x+\hat{\mu})_b-\Psi(x+\hat{\mu})^*_a\Phi_{\mu}(x)^*_{ab}\Psi(x)_b\right]\ ,
\end{aligned}
\end{equation}
where matrix elements of disorder fields, i.e. $\Phi_\mu(x)_{ab}$, $a,b=1,2,\ldots,N$ are distributed according to density:
\begin{equation}\label{gaussian_measure}
f\left[\Phi_\mu(x)_{ab}\right]=Ce^{-N\abs{\Phi_\mu(x)_{ab}}^2}
\end{equation}
with $C$ being a normalization constant. The theory may be formally described by the same Hamiltonian kernel $h$ as in subsection (\ref{hamilton_operator}) with SU(N) fields substituted by Gaussian fields. In terms of the Grassmann valued fermion field $\psi(x,t)$, with $t$ labeling points in the extra dimension, the model is defined by the action:
\begin{equation}
\begin{aligned}
{\cal I}=&-N\sum_{x,a,b,\mu}\abs{\Phi_\mu(x)_{ab}}^2+\sum_{x,t,a}\bar{\psi}_a(x,t)[1+\hat{\gamma}_5(x)\hat{\partial_t}]\psi(x,t)_a\\
+&\kappa\sum_{x,t,a,b,\mu}\eta_\mu(x)\left[\bar{\psi}(x,t)_a\Phi_{\mu}(x)_{ab}\psi(x+\hat{\mu},t)_b-\bar{\psi}(x+\hat{\mu},t)_a\Phi_{\mu}(x)^*_{ab}\psi(x,t)_b\right]\ ,
\end{aligned}
\end{equation}
whereas the partition function of the theory is:
$$
Z=\int\prod_{x,\mu,a,b}d\Phi_{\mu}(x)_{ab}d~\overline{\Phi_{\mu}(x)_{ab}}~\prod_{x,a}d\psi(x)_ad\bar\psi(x)_a~e^{\cal I}\ .
$$
Gaussian fields can be integrated using the formula:
$$
\int dzd\bar{z}~e^{-\alpha\abs{z}^2+\bar\beta z+\gamma\bar{z}}=\frac{2\pi i}{\alpha}~e^{\bar\beta\gamma/\alpha}\ ,
$$
where $\alpha$ is a positive real number and $\beta,\gamma$ are complex numbers. The effective action of the theory which remains after integration of Gaussian fields is:
\begin{equation}
\begin{aligned}
S&=\sum_{x,t,a}\bar{\psi}_a(x,t)[1+\hat{\gamma}_5(x)\hat{\partial_t}]\psi(x,t)_a\\
&+\frac{\kappa^2}{N}\sum_{x,\mu,t,t',a,b}\bar{\psi}(x,t)_b\psi(x,t')_b\bar{\psi}(x+\hat{\mu},t')_a\psi(x+\hat{\mu},t)_a\ .
\end{aligned}
\end{equation}
This is precisely equation (\ref{psi_action}), which is used as an approximation of the full fermion theory (\ref{pure_fermion}) in the case of SU(N) disorder. This shows that SU(N) and Gaussian disorder theories are closely related. In the following subsection we elaborate further this relationship.

\subsection{Embedded gauge fields}

In this subsection we identify embedded gauge fields within Gaussian fields and show that the large $N_t$ theory is effectively a local Yang-Mills theory. We begin by giving the expression of the effective action which remains after fermions are integrated out. Following the same steps as in subsection (\ref{massive_fermions}) and taking into the account the Gaussian measure (\ref{gaussian_measure}) the effective action of the theory is:
\begin{equation}\label{eff_action_Phi}
S_{\text{eff}}(\Phi)=N\sum_{x,a,b,\mu}\abs{\Phi_\mu(x)_{ab}}^2-\frac{N_t}{2}~\text{Tr~}\sqrt{\1+\kappa^2h_{o,\Phi}^2}
\end{equation}
with:
\begin{equation}
h_{o,\Phi}=\hat{\gamma}_5\sum_\mu\eta_\mu(\Phi_\mu-\Phi_\mu^*)\ ,~~~~~~~~(\Phi_\mu)_{xyab}=\Phi_\mu(x)_{ab}\delta_{x+\hat\mu,y}\ .
\end{equation}
Note that the trace in the right hand side of (\ref{eff_action_Phi}) is taken in the tensor product space of the lattice sites and $N\times N$ matrices. Using the polar decomposition of Gaussian fields:
\begin{equation}
\Phi_\mu(x)=\phi_\mu(x)U_\mu(x)\ ,
\end{equation}
gauge fields are identified by the U(N) factor $U_\mu(x)$, where $\phi_\mu(x)$ are positive definite Hermitian matrices. In the following we keep the gauge fields fixed and find the saddle point action in the large $N_t$ limit using the solution Ansatz:
\begin{equation}
\phi_\mu(x)_{ab}=\varphi~\delta_{ab}\ ,
\end{equation}
where $\varphi$ is a real value. A more general Ansatz would include terms which give $O(1/N_t)$ contributions to the saddle point action. Since we are interested in the leading contributions to the effective action we stay with the above solution Ansatz. Note also that both $N$ and $N_t$ are large and we relate them by the $\kappa$ dependent factor:
\begin{equation}
r(\kappa)=\frac{N}{N_t}\ .
\end{equation}
Using this definition and the solution Ansatz the action is a function of a single variable $\varphi$:
\begin{equation}\label{varphi_action}
S(\varphi)/N_t=\varphi^2r(\kappa)VNd-\frac{1}{2}~\text{Tr~}\sqrt{\1+\varphi^2\kappa^2h_o^2}\ ,
\end{equation}
where $h_o$ is the fermion matrix in the background of the gauge field $U_\mu(x)$. The saddle point equation $S'(\varphi)=0$ yields the nontrivial solution $\varphi_o$ given implicitly by the equation:
\begin{equation}\label{saddle_point_equation}
r(\kappa)=\frac{1}{4VNd}~\text{Tr~}\frac{\kappa^2h_o^2}{\sqrt{\1+\varphi_o^2\kappa^2h_o^2}}\ .
\end{equation}
Note that the solution $\varphi_o$ is gauge field dependent. As shown in the next subsection, an explicit solution may be computed in the limit of vanishing $\kappa$. Alternatively, one may assign to the action (\ref{varphi_action}) an approximate as well as gauge field independent solution. Such a solution corresponds to an approximate saddle point. This way, one may proceed as in subsection (\ref{massive_fermions}) and obtain a local theory of Yang-Mills fields. Therefore, Gaussian and U(N) disorder theories are related at the approximate saddle point of the Gaussian disordered action in the large $N_t$ limit. In the next subsection we give an example of a precise relationship.

\subsection{Saddle point Yang-Mills theory}

In this subsection we compute the saddle point action of the theory with Gaussian disorder fields in the limit of vanishing $\kappa$. Our starting point is the saddle point equation (\ref{saddle_point_equation}). Making the following Ansatz for the left hand side:
\begin{equation}\label{r_kappa_ansatz}
r(\kappa)=a_2\kappa^2+a_4\kappa^4\ ,
\end{equation}
with $a_2,a_4$ being real constants, and matching it to the right hand side expansion of (\ref{saddle_point_equation}) in $\kappa$ we find:
\begin{equation}
a_2=\frac{\text{Tr~}h_o^2}{4VNd}\ ,~~~~~~~~~~~~\varphi_o^2=-\frac{8a_4VNd}{\text{Tr~}h_o^4}\ .
\end{equation}
Note that we have neglected the higher powers of the expansion of the right hand side since we seek the limit of vanishing $\kappa$. We have also the freedom to select a small value of $a_4$ in order to control the matching error of the expansion. Substitution of $a_2$ and $\varphi_o^2$ to the action (\ref{varphi_action}) give the effective theory of Yang-Mills fields:
\begin{equation}
S_{\text{eff}}(U)/N_t\propto-\frac{4a_4^2(VNd)^2\kappa^4}{\text{Tr~}h_o^4}+O(\kappa^6)\ .
\end{equation}
Since the leading term does not look like the standard plaquette action of Wilson we expand $\text{Tr~}h_o^4$ in terms of gauge fields and get:
\begin{equation}
S_{\text{eff}}(U)/N_t\propto-\frac{4a_4^2(VNd)^2\kappa^4}{6VNd+4\sum_{\mu\neq\nu}\text{Tr~}(\1-U_\mu U_\nu U_\mu^* U_\nu^*)}+O(\kappa^6)\ .
\end{equation}
For smooth gauge fields close to continuum limit, the plaquette terms $\text{Tr}U_\mu U_\nu U_\mu^* U_\nu^*$ are close to one. This property allows us to write the right hand side as a geometric series of $4\sum_{\mu\neq\nu}\text{Tr}(\1-U_\mu U_\nu U_\mu^* U_\nu^*)/6VNd$ with the leading term:
\begin{equation}
S_{\text{eff}}(U)/N_t\propto\frac{4}{9}a_4^2\kappa^4\sum_{\mu\neq\nu}\text{Tr~}U_\mu U_\nu U_\mu^* U_\nu^*+O(\kappa^6)\ .
\end{equation}
If we insist to maintain the scaling of $N_t$ to $\kappa$ as defined in (\ref{N_t_kappa}) then the leading terms of this theory and the one of (\ref{ym2}) are the same provided we select $a_4=3/4$. This calculation shows that close to continuum limit the saddle point action of the theory with Gaussian fields yields a similar Yang-Mills leading term as in the case of the theory with SU(N) disorder fields. Note that the form of the $r(\kappa)$ Ansatz (\ref{r_kappa_ansatz}) is crucial to arrive to this conclusion. If there is no relation between $N$ and $N_t$ the theory may not be local.

\section{Group integration}
\label{group_integration}

This section is written to make the paper self contained. We begin first with some integration rules in the unitary groups.

\subsection{Group integration rules}
\label{group_integration_rules}

Unitary group integration rules used in this paper rely on the Haar measure. These rules are known and we point the reader to the paper of Creutz, reference \cite{creutz_group_integration},\footnote{There is a slight difference with our formulas since this reference uses the group $SU(N)$.} for a detailed account. Here we would like to give a few useful results. For example, invariant group integration arguments lead to the conclusion:
\begin{equation}
\int dU~U_{ab}=0\ .
\end{equation}
One can extend this result for the product of $n$ matrix elements as long as $n\neq N$. In the case $n=N$ the integral is nonvanishing if indices lead to a group invariant quantity such as the determinant of $U$,\footnote{In case of $SU(N)$ group the determinant is one.} i.e.:
\begin{equation}\label{det_U}
\int dU~U_{1a_1}U_{2a_2}\cdots U_{Na_N}=\frac{\det U}{N!}\epsilon_{a_1a_2\ldots a_N}\ ,
\end{equation}
where $a_1,a_2,\ldots,a_N$ is a permutation of $1,2,\ldots,N$ and $\epsilon_{a_1a_2\ldots a_N}$ is the rank $N$ totally antisymmetric tensor. Indeed, if we multiply both sides by $\epsilon_{a_1a_2\ldots a_N}$ and sum over all permumations $a_1,a_2,\ldots,a_N$ we get an identity. Another useful integral is:
\begin{equation}\label{group_integral}
\int dU~U_{ab}{U^*}_{cd}=\frac{1}{N}\delta_{ad}\delta_{bc}\ .
\end{equation}
One can be convinced about the normalization and the $\delta_{ad}$ factor by taking $b=c$ and summing both sides over $b$. The same argument justifies the factor $\delta_{bc}$. This integral is also derived at the end of the next subsection applying the results obtained therein.

\subsection{Computation of one-link integrals}

In this subsection we deal with the computation of one-link integrals of the type:
\begin{equation}
e^{W(\bar\psi,\psi,\bar\chi,\chi)}=\int dU~e^{\sum_{a,b}\left(\sum_t\bar{\psi}_a^tU_{ab}\chi_b^t+\sum_{t'}\bar{\chi}_a^{t'}U^*_{ab}\psi_b^{t'}\right)}\ ,
\end{equation}
where $a,b$ indices run from $1$ to $N$, the $t$ index runs from $1$ to $N_t$, the trace is taken in $t$-space and $dU$ is the Haar measure of the $U(N)$ group. Using invariance properties of the Haar measure, the integral depends only on gauge invariant quantities:
\begin{equation}
\sum_a\bar{\psi}^t_a\psi^{t'}_a\ ,~~~~~~~~\sum_a\bar{\chi}^{t'}_a\chi^t_a\ .
\end{equation}
On the other hand, the integral depends also on bilinear Grassmann sums:
\begin{equation}
\sum_t\bar{\psi}^t_a\chi^t_b\ ,~~~~~~~~\sum_{t'}\bar{\chi}^{t'}_a\psi^{t'}_b
\end{equation}
since they are invariant with respect to invertible $N_t\times N_t$ matrix transformations. Therefore, if we define the $N_t\times N_t$ matrix:
\begin{equation}\label{Lambda}
\Lambda(t,t'')=\frac{1}{N^2}\sum_{t',a,b}\bar{\psi}_b^t\psi_b^{t'}\bar{\chi}_a^{t'}\chi_a^{t''}\ ,
\end{equation}
the integral is a function of $t$-space traces of this matrix:
\begin{equation}\label{invariant_traces}
\text{tr~}\Lambda\ ,\text{tr~}\Lambda^2\ ,\ldots,\text{tr~}\Lambda^{N_t}\ ,
\end{equation}
i.e. it can be written in the form:
\begin{equation}\label{integral}
e^{N~\text{tr~}F(\Lambda)}=\int dU~e^{\sum_{a,b}\left(\sum_t\bar{\psi}_a^tU_{ab}\chi_b^t+\sum_{t'}\bar{\chi}_a^{t'}U^*_{ab}\psi_b^{t'}\right)}\ ,
\end{equation}
where $F$ is a matrix valued function defined by its power series expansion. Taking the derivative of both sides with respect to $\bar\psi^{t_1}_b$ and then $\psi^{t}_b$ in this order, multiplying by $\bar\psi^{t_2}_a\psi^t_a$, summing over $a,b$ and $t$ we find:
\begin{equation}
\sum_{a,b,t}\bar\psi^{t_2}_a\psi^t_a\frac{\partial^2e^{N~\text{tr~}F(\Lambda)}}{\partial\psi^{t}_b\partial\bar\psi^{t_1}_b}=-N^2\Lambda_{t_2t_1}e^{N~\text{tr~}F(\Lambda)}\ .
\end{equation}
The left hand side may be written in terms of derivatives of $\text{tr~}F(\Lambda)$ with respect to matrix elements of $\Lambda$. This way, we obtain a system of $N_t^2$ coupled second order differential equations for $\text{tr}F(\Lambda)$:
\begin{equation}
\begin{aligned}
&\sum_{t't''t_3t_4}\left(N^2\frac{\partial~\text{tr~}F}{\partial\Lambda_{t't''}}\frac{\partial~\text{tr~}F}{\partial\Lambda_{t_3t_4}}+N\frac{\partial^2~\text{tr~}F}{\partial\Lambda_{t't''}\partial\Lambda_{t_3t_4}}\right)\sum_{a,b,t}\bar\psi^{t_2}_a\psi^t_a\frac{\partial\Lambda_{t't''}}{\partial\psi^{t}_b}\frac{\partial\Lambda_{t_3t_4}}{\partial\bar\psi^{t_1}_b}\\
&+\sum_{t_2t_4}\frac{\partial~\text{tr~}F}{\partial\Lambda_{t_3t_4}}\sum_{a,b,t}\bar\psi^{t_2}_a\psi^t_a\frac{\partial^2\Lambda_{t_3t_4}}{\partial\psi^{t}_b\partial\bar\psi^{t_1}_b}=-N^2\Lambda_{t_2t_1}\ ,
\end{aligned}
\end{equation}
supplemented by suitable boundary conditions. For example, the set:
\begin{equation}\label{conditions}
\left.F(\Lambda)_{t_2t_1}\right|_{\Lambda=0}=0\ ,~~~~~~~~\left.\frac{\partial~\text{tr~}F}{\partial\Lambda_{t_1t_2}}\right|_{\Lambda=0}=-\delta_{t_2t_1}
\end{equation}
guarantees that the solution is regular around zero. Using the definition of $\Lambda$, equation (\ref{Lambda}) the system to be solved is:
\begin{equation}
\begin{aligned}
&\sum_{t_4}\Lambda_{t_2t_4}\frac{\partial~\text{tr~}F}{\partial\Lambda_{t_1t_4}}-\sum_{t'}\left(\sum_{t4}\Lambda_{t't_4}\frac{\partial~\text{tr~}F}{\partial\Lambda_{t_1t_4}}\right)\left(\sum_{t''}\Lambda_{t_2t''}\frac{\partial~\text{tr~}F}{\partial\Lambda_{t't''}}\right)\\
&-\frac{1}{N}\sum_{t'}\left(\sum_{t''t_4}\Lambda_{t_2t''}\Lambda_{t't_4}\frac{\partial^2~\text{tr~}F}{\partial\Lambda_{t't''}\partial\Lambda_{t_1t_4}}\right)+\Lambda_{t_2t_1}=0\ .
\end{aligned}
\end{equation}
This system may be written in matrix notations in the form:
\begin{equation}\label{Lambda_form}
\Lambda\frac{\partial~\text{tr}F}{\partial\Lambda}-\left(\Lambda\frac{\partial~\text{tr}F}{\partial\Lambda}\right)^2-\frac{1}{N}\left(\Lambda\frac{\partial}{\partial\Lambda}\right)^2\text{tr~}F+\Lambda=0\ .
\end{equation}
Since we are interested in the large $N$ limit we drop the third term and find:
\begin{equation}\label{differential_F}
\left(\Lambda\frac{\partial~\text{tr}F}{\partial\Lambda}\right)^2-\Lambda\frac{\partial~\text{tr}F}{\partial\Lambda}-\Lambda=0\ .
\end{equation}
This is an algebraic quadratic matrix equation for $\Lambda\frac{\partial~\text{tr}F}{\partial\Lambda}$. Its solution poses no problem and is given by:
\begin{equation}\label{first_order_diff_eq}
\Lambda\frac{\partial~\text{tr}F}{\partial\Lambda}=\frac{1-\sqrt{1+4\Lambda}}{2}\ ,
\end{equation}
where the second condition in equations (\ref{conditions}) is taken into account and the matrix square root is defined in terms of its power series expansion. A solution for the matrix $F(\Lambda)$ that satisfies this equation as well as the condition $F(0)=0$ is:
\begin{equation}\label{one-link_result}
F(\Lambda)=1-\sqrt{1+4\Lambda}+\ln\frac{1+\sqrt{1+4\Lambda}}{2}\ .
\end{equation}
In this case too, the matrix logarithm is defined in terms of its power series expansion.\footnote{Note that boundary conditions specified in (\ref{conditions}) apply in the case when $\text{tr}\Lambda$ is small. In this case the first term in the left hand side of the equation (\ref{differential_F}) may be neglected and one finds $F(\Lambda)=-\Lambda$.}

As an application let us derive the group integral (\ref{group_integral}). Taking the fourth derivative of both sides of the integral (\ref{integral}) with respect to $\bar\psi,\psi,\bar\chi,\chi$ Grassmann variables in this order one finds:
\begin{equation}
\left.\frac{\partial^4e^{N~\text{tr~}F(\Lambda)}}{\partial\chi^{t}_{b_2}\partial\bar\chi^{t'}_{b_1}\partial\psi^{t'}_{a_1}\partial\bar\psi^{t}_{a_2}}\right|_{\psi=\bar\psi=\chi=\bar\chi=0}=-\int dU~U_{a_1b_2}{U^*}_{b_1a_2}\ .
\end{equation}
Substituting $F(\Lambda)$ in the left hand side using the result, equation (\ref{one-link_result}), one finds:
\begin{equation}
\left.\frac{\partial^4e^{N~\text{tr~}F(\Lambda)}}{\partial\chi^{t}_{b_2}\partial\bar\chi^{t'}_{b_1}\partial\psi^{t'}_{a_1}\partial\bar\psi^{t}_{a_2}}\right|_{\psi=\bar\psi=\chi=\bar\chi=0}=-\frac{1}{N}\delta_{a_1a_2}\delta_{b_1b_2}\ .
\end{equation}
Comparing right hand sides of last two equations, equation (\ref{group_integral}) is thus established. Other interesting integrals can be computed using this method.

\subsection{Relation to other work}

Our derivation can be related to the one of reference \cite{kluberg-stern_et_al}. In this reference the derivative of $\text{tr~}F$ is taken with respect to invariant traces:
\begin{equation}
\lambda_1=\text{tr~}\Lambda\ ,~~~~\lambda_2=\text{tr~}\Lambda^2\ ,~~~~\ldots,~~~~\lambda_{N_t}=\text{tr~}\Lambda^{N_t}\ .
\end{equation}
The resulting system of differential equations is related to ours, equation (\ref{Lambda_form}), if the derivative of $\text{tr~}F$ with respect to matrix elements of $\Lambda$ is defined by the expression:
$$
\frac{\partial~\text{tr}F}{\partial\Lambda}=\Lambda\frac{\partial~\text{tr}F}{\partial\lambda_1}+2\Lambda\frac{\partial~\text{tr}F}{\partial\lambda_2}+\cdots+N_t\Lambda\frac{\partial~\text{tr}F}{\partial\lambda_{N_t}}\ .
$$
In order to find the solution, reference \cite{kluberg-stern_et_al} diagonalizes the matrix $\Lambda$, whereas reference \cite{kawamoto_smit} relies on the "strong coupling" solution of Brezin and Gross \cite{brezin_gross}, their solution being found also by diagonalizing $\Lambda$.

\section{QCD at strong coupling}
\label{qcd_strong}

The large $N$ limit of QCD at strong coupling has been studied in the past. Notable references are Kluberg-Stern {\it et. al.} \cite{kluberg-stern_et_al} and Kawamoto and Smit \cite{kawamoto_smit}. The aim of this appendix is to derive the main results of QCD at strong coupling using the bosonization approach employed in this paper. The action is given by equation (\ref{I_action}) but without time derivative term, i.e.:
\begin{equation}
\begin{aligned}
I=&m_f\sum_{x,t,a}\bar{\psi}_a(x,t)\psi(x,t)_a\\
+&\kappa\sum_{x,t,a,b,\mu}\eta_\mu(x)\left[\bar{\psi}(x,t)_aU_{\mu}(x)_{ab}\psi(x+\hat{\mu},t)_b-\bar{\psi}(x+\hat{\mu},t)_aU_{\mu}(x)^*_{ab}\psi(x,t)_b\right]\ ,
\end{aligned}
\end{equation}
where $\kappa$ is now fixed at the value of $1/2$. Everything else being the same, integrating gauge fields as in subsection \ref{integration_gauge_fields} we get:
\begin{equation}\label{fermion_theory}
S=m_f\sum_{x,t,a}\bar{\psi}_a(x,t)\psi(x,t)_a+N\sum_{x,\mu,t}F\left[-\frac{\kappa^2}{N^2}\sum_{t',a,b}\bar{\psi}(x,t)_b\psi(x,t')_b\bar{\psi}(x+\hat{\mu},t')_a\psi(x+\hat{\mu},t)_a\right]\ ,
\end{equation}
where $F(.)$ is defined by the expression (see (\ref{one-link_result})):
\begin{equation}\label{F_function1}
F(\Lambda)=1-(1+4\Lambda)^{\frac{1}{2}}+\ln\frac{1+(1+4\Lambda)^{\frac{1}{2}}}{2}\ .
\end{equation}
In this paper we approximate the right hand side with the leading order result $F(\Lambda)=-\Lambda+O(\Lambda^2)$. It is this approximation that will be tested in the case of strong coupling QCD. Following the same steps as in section \ref{matrix_theory} the final expression of the bosonic effective action is (see equation (\ref{sigma_action})):
\begin{equation}\label{Sigma1_action}
S_\Sigma=N\sum_{x,t}\left\{\ln[m_f+\Sigma(x)]\right\}(t,t)-\frac{N}{2\kappa^2}\sum_{x,y,t,t'}\Sigma(x,t,t')(A^{-1})(x,y)\Sigma(y,t',t)\ .
\end{equation}

\subsection{Large $N$ solution}

For future references on QCD at strong coupling we make this section self contained. Therefore, a few steps of the large $N$ solution of section \ref{solution1} will be repeated here. In order to find the large $N$ solution we find first the field that makes the action stationary and then compute an effective action by computing fluctuations around such a solution. The stationary field should satisfy the necessary first order conditions, which in our case is the system of equations:
\begin{equation}
\frac{\partial S_\Sigma}{\partial\Sigma(x,t,t')}=0\ ,~~~~x\in\Lambda_d\ ,~~~~t,t'=1,2\ldots,N_t\ ,
\end{equation}
where $S_\Sigma$ is the bosonic action, equation (\ref{Sigma1_action}). The derivative of the first term is taken by expanding the matrix logarithm as a power series on $\Sigma$. This way, we obtain the system of equations:
\begin{equation}\label{Sigma2_eq}
\left\{\frac{1}{m_f+\Sigma(x)}\right\}(t,t')=\frac{1}{\kappa^2}\sum_y(A^{-1})(x,y)\Sigma(y,t,t')\ .
\end{equation}
Denoting by $G(x)=1/[m_f+\Sigma(x)]$ the $N_t\times N_t$ matrix, equations take the form:
\begin{equation}\label{G2_eq}
\left\{\frac{1}{G(x)}\right\}(t,t')=m_f\delta_{t,t'}+\kappa^2\sum_{\mu}[G(x+\hat{\mu},t,t')+G(x-\hat{\mu},t,t')]\ .
\end{equation}
This is a system of matrix valued quadratic equations which is solved using the Ansatz:
\begin{equation}\label{Ansatz1}
G(x,t,t')=G_o\delta_{t,t'}+\tilde{G}(x,t,t')\ ,
\end{equation}
where $\tilde{G}(x,t,t')$ is a small fluctuation field around the uniform solution $G_o\delta_{t,t'}$. Expanding the left hand side of (\ref{G2_eq}) up to the first order in power series of the matrix $G_o^{-1}\tilde{G}(x)$, we get a quadratic equation for the matrix $G_o$:
\begin{equation}
\frac{1}{G_o}-m_f-2d\kappa^2G_o=0
\end{equation}
with the solution:
\begin{equation}\label{G1_o}
G_o=\frac{-m_f+\sqrt{m_f^2+8d\kappa^2}}{4d\kappa^2}\ ,
\end{equation}
as well as the first order constraint in the fluctuating field $\tilde{G}(x,t,t')$:
\begin{equation}\label{constraint1}
\tilde{G}(x,t,t')+\kappa^2G_o^2\sum_{\mu}[\tilde{G}(x+\hat{\mu},t,t')+\tilde{G}(x-\hat{\mu},t,t')]=0\ .
\end{equation}
Substituting the Ansatz, equation (\ref{Ansatz1}), into the effective action (\ref{Sigma1_action}), using equations (\ref{Sigma2_eq}), (\ref{G2_eq}) and the first order constraint (\ref{constraint1}), as well as expanding the logarithm up to the second order in powers of the $G_o^{-1}G(x)$ matrix we get:
\begin{equation}\label{solution}
\begin{aligned}
S_{\Sigma}/N=&-N_tV\left(\ln G_o+d\kappa^2 G_o^2\right)\\
+&\frac{\kappa^2}{2}\left(\frac{1}{\kappa^2G_o^2}-2d\right)\sum_{x,t,t'}\tilde{G}(x,t,t')\tilde{G}(x,t',t)\\
+&\frac{\kappa^2}{2}\sum_{x,t,t',\mu}\left[\tilde{G}(x+\hat{\mu},t,t')-\tilde{G}(x,t,t')\right]\left[\tilde{G}(x+\hat{\mu},t',t)-\tilde{G}(x,t',t)\right]\\
+&O\left[G_o^{-3}\tilde{G}(x)^3\right]\ ,
\end{aligned}
\end{equation}
where the translation invariance on the lattice, i.e. the identity $\sum_x\tilde{G}(x+\hat{\mu},t,t')^2=\sum_x\tilde{G}(x,t,t')^2$ has also been used. From this expression one can infer the free energy of the theory:
\begin{equation}\label{free_energy1}
{\cal F}=NN_tV\left(\ln G_o+d\kappa^2 G_o^2\right)\ .
\end{equation}
On the other hand, equation (\ref{solution}) can be written in the form:
\begin{equation}\label{effective_theory1}
-S_{\tilde{G}}/(\kappa^2 N)=M^2\sum_{x}\text{tr~}e^{i\tilde{G}(x)}+\sum_{x,\mu}\text{tr~}e^{i\tilde{G}(x+\hat{\mu})}e^{-i\tilde{G}(x)}+O\left[G_o^{-3}\tilde{G}(x)^3\right]+h.c.\ ,
\end{equation}
with the trace taken in $t$-space and where we have denoted:
\begin{equation}\label{M2}
M^2=\frac{1}{\kappa^2G_o^2}-2d=\frac{m_f^2+m_f\sqrt{m_f^2+8d\kappa^2}}{2\kappa^2}\ .
\end{equation}
Neglecting cubic order corrections, the resulting effective theory (\ref{effective_theory1}) is a non-linear sigma model of massive bosons with mass $M$ given by equation (\ref{M2}). For vanishing $m_f$, the mass squared vanishes linearly with $m_f$:
\begin{equation}
M^2=\frac{m_f}{\kappa}\sqrt{2d}+O(m_f^2)\ .
\end{equation}
For vanishing quark mass the theory has an exact global $U(N_t)_L\times U(N_t)_R$ symmetry, i.e. the effective action is invariant with respect to global transformations:
\begin{equation}\label{chiral_symmetry1}
e^{i\tilde{G}(x+\hat{\mu})}\rightarrow Ue^{i\tilde{G}(x+\hat{\mu})}V^*\ ,~~~~~~~~e^{i\tilde{G}(x)}\rightarrow Ve^{i\tilde{G}(x)}U^*\ .
\end{equation}
The chiral condensate of the theory does not vanish and therefore the chiral symmetry (\ref{chiral_symmetry1}) is spontaneously broken to $U(N_t)$. In order to see this, we compute the chiral condensate, which is defined by expressions:
\begin{equation}
\zeta=\lim_{m_f\rightarrow 0}\lim_{V\rightarrow\infty}\frac{1}{V}\sum_{x,t,a}\left\langle \bar{\psi}(x,t)_a\psi(x,t)_a \right\rangle=\lim_{m_f\rightarrow 0}\lim_{V\rightarrow\infty}\frac{1}{V}\frac{\partial \ln Z}{\partial m_f}=-\lim_{m_f\rightarrow 0}\lim_{V\rightarrow\infty}\frac{1}{V}\frac{\partial {\cal F}}{\partial m_f}\ .
\end{equation}
Using the result (\ref{free_energy1}) for the free energy and substituting the solution $G_o$, equation (\ref{G1_o}), we get:
\begin{equation}\label{condensate1}
\zeta=\left.\frac{NN_t}{\sqrt{2d\kappa^2}}\right|_{\kappa=\frac{1}{2}}=NN_t\sqrt{\frac{2}{d}}\ .
\end{equation}
Therefore, the chiral symmetry of QCD at strong coupling is broken spontaneously to $U(N_t)$ group, whereas the spectrum has $N_t^2$ Goldstone bosons.

We note that our results rely on the leading order approximation $F(\Lambda)=-\Lambda+O(\Lambda^2)$. Our condensate and boson mass differ with the exact results of references \cite{kluberg-stern_et_al,kawamoto_smit} by a $O(1)$ prefactor at vanishing fermion mass. This small difference does not alter the physical picture and thus justifies our approximation. In order to make these differences transparent, we summarize below the approach and results obtained in these references.

\subsection{Relation to other work}

If we want our results to be exact, we may follow the bosonization approach of references \cite{kluberg-stern_et_al,kawamoto_smit}. Another possibility is to use the Lagrange multiplier method as shown in a series of papers related to the SYK model, see for example Gross and Rosenhaus \cite{gross_rosenhaus}, as well as  Kitaev and Suh \cite{kitaev_suh}.

Kluberg-Stern {\it et. al.} paper starting point is the observation that integration of functions which depend on Grassmann pairs $\bar\eta^t\eta^{t'},t,t'=1,\ldots,k$ can be cast in the form of the following matrix determinant:
\begin{equation}
\int\prod_{t=1}^kd\eta^td\bar\eta^t~f(\bar\eta^1\eta^1,\bar\eta^1\eta^2,\ldots,\bar\eta^k\eta^k)=\left|
\begin{matrix}
\frac{\partial f}{\partial(\bar\eta^1\eta^1)} & & & \frac{\partial f}{\partial(\bar\eta^1\eta^k)}\\
\vdots\\
\frac{\partial f}{(\partial\bar\eta^k\eta^1)} & & & \frac{\partial f}{\partial(\bar\eta^k\eta^k)}
\end{matrix}
\right|_{\eta=\bar\eta=0}\ .
\end{equation}
The proof can be established by noting that only the order $k$ term of Grassmann pairs power series expansion of $f$ contributes in the integral. If Grassmann pairs of the right hand side are formally substituted by real valued matrix elements $\sigma_{tt'}=0,t,t'=1,\ldots,k$ we get the identity:
\begin{equation}\label{berezin_integral}
\int\prod_{t=1}^kd\eta^td\bar\eta^t~f(\bar\eta^1\eta^1,\bar\eta^1\eta^2,\ldots,\bar\eta^k\eta^k)=\left|
\begin{matrix}
\frac{\partial f}{\partial\sigma_{11}} & & & \frac{\partial f}{\partial\sigma_{1k}}\\
\vdots\\
\frac{\partial f}{\partial\sigma_{k1}} & & & \frac{\partial f}{\partial\sigma_{kk}}
\end{matrix}
\right|_{\sigma=0}\ .
\end{equation}
Substituting the Fourier representation:
\begin{equation}
f(\sigma)=\frac{1}{(2\pi)^{k^2}}\int\prod_{tt'}d\lambda_{tt'}~e^{-i\sum_{tt'}\sigma_{tt'}\lambda_{t't}}\tilde{f}(\lambda)
\end{equation}
in the right hand side we get:
\begin{equation}
\int\prod_{t=1}^kd\eta^td\bar\eta^t~f(\bar\eta^1\eta^1,\bar\eta^1\eta^2,\ldots,\bar\eta^k\eta^k)=\frac{(-i)^k}{(2\pi)^{k^2}}\int\prod_{tt'}d\lambda_{tt'}\det\lambda~\tilde{f}(\lambda)\ .
\end{equation}
The Fourier transformed function $\tilde{f}$ can be written in terms of the original function using the inverse Fourier transform and the final expression is:
\begin{equation}
\int\prod_{t=1}^kd\eta^td\bar\eta^t~f(\bar\eta^1\eta^1,\bar\eta^1\eta^2,\ldots,\bar\eta^k\eta^k)=\frac{(-i)^k}{(2\pi)^{k^2}}\int\prod_{tt'}d\lambda_{tt'}d\sigma_{tt'}~\det\lambda~e^{i\sum_{tt'}\sigma_{tt'}\lambda_{t't}}f(\sigma)\ .
\end{equation}
If $f$ depends on Grassmann sums $\sum_{a=1}^N\bar\eta_a^t\eta_a^{t'},t,t'=1,\ldots,k$, as it is the case in our application, the Berezin integral, equation (\ref{berezin_integral}), gives:
\begin{equation}\label{bosonization}
\int\prod_{a=1}^N\prod_{t=1}^kd\eta_a^td\bar\eta_a^t~f\left(\sum_{a=1}^N\bar\eta_a^1\eta_a^1,\sum_{a=1}^N\bar\eta_a^1\eta_a^2,\ldots,\sum_{a=1}^N\bar\eta_a^k\eta_a^k\right)=\left|
\begin{matrix}
\frac{\partial}{\partial\sigma_{11}} & & & \frac{\partial}{\partial\sigma_{1k}}\\
\vdots\\
\frac{\partial}{\partial\sigma_{k1}} & & & \frac{\partial}{\partial\sigma_{kk}}
\end{matrix}
\right|^N\left.f(\sigma)\right|_{\sigma=0}\ ,
\end{equation}
whereas the result reads:
\begin{equation}
\begin{aligned}
&\int\prod_{a=1}^N\prod_{t=1}^kd\eta_a^td\bar\eta_a^t~f\left(\sum_{a=1}^N\bar\eta_a^1\eta_a^1,\sum_{a=1}^N\bar\eta_a^1\eta_a^2,\ldots,\sum_{a=1}^N\bar\eta_a^k\eta_a^k\right)\\
=&~~~~\frac{(-i)^k}{(2\pi)^{k^2}}\int\prod_{tt'}d\lambda_{tt'}d\sigma_{tt'}~\det\lambda^Ne^{i\sum_{tt'}\sigma_{tt'}\lambda_{t't}}f(\sigma)\ ,
\end{aligned}
\end{equation}
This way, the fermion theory can be cast in the form of the bosonic theory with an action $S$ which depends on matrix elements $\lambda_{tt'},\sigma_{tt'},t,t'=1,2,\ldots,k$:
\begin{equation}
-S(\lambda,\sigma)=\text{tr~}(N\ln\lambda+i\lambda^T\sigma)+\ln f(\sigma)\ .
\end{equation}
Using this technique and adopting the normalization of fermion bilinears as in reference \cite{kluberg-stern_et_al}, i.e. $\bar\psi\psi\rightarrow i\bar\psi\psi$, one can show that the large $N$ solution of the fermion theory (\ref{fermion_theory}) is given at the saddle point $\sigma_+\delta_{tt'}$ with:
\begin{equation}
i\sigma_+=\left.\frac{-m_f\left(1-\frac{1}{d}\right)+\sqrt{m_f^2+2d-1}}{d\left(1+\frac{m_f^2}{d^2}\right)}\right|_{m_f=0}=\frac{\sqrt{2d-1}}{d}\ ,
\end{equation}
whereas the chiral condensate is:
\begin{equation}
\zeta=NN_t\sqrt{\frac{2}{d}}\sqrt{1-\frac{1}{2d}}\ .
\end{equation}

Now let us turn to the approach followed by Kawamoto and Smit \cite{kawamoto_smit}. These authors use the bosonization of the fermion determinant based on the identity:
\begin{equation}\label{det_J}
\int dU~\left(\det RU\right)^{-N}e^{\text{~tr}JRU}=c_N^k\left(\det J\right)^N\ ,
\end{equation}
where $dU$ is the Haar measure on the $U(k)$ group, $R$ is a Hermitian and positive definite $k\times k$ matrix and $c_N^k$ are defined by the expression:
\begin{equation}
c_N^k=\frac{0!~1!\cdots(k-1)!}{N!~(N+1)!\cdots(N+k-1)!}\ .
\end{equation}
Since the left hand side of the equation is independent of $R$ one can define $M=RU$ and combine equations (\ref{det_J}), (\ref{bosonization}) to get:
\begin{equation}
\begin{aligned}
&\int\prod_{a=1}^N\prod_{t=1}^kd\eta_a^td\bar\eta_a^t~f\left(\sum_{a=1}^N\bar\eta_a^1\eta_a^1,\sum_{a=1}^N\bar\eta_a^1\eta_a^2,\ldots,\sum_{a=1}^N\bar\eta_a^k\eta_a^k\right)\\
&=\frac{1}{c_N^k}\int dU~e^{-N\text{~tr}\ln M}~e^{\text{~tr}
\begin{pmatrix}
\frac{\partial}{\partial\sigma_{11}} & & & \frac{\partial}{\partial\sigma_{1k}}\\
\vdots\\
\frac{\partial}{\partial\sigma_{k1}} & & & \frac{\partial}{\partial\sigma_{kk}}
\end{pmatrix}
M}
\left.f(\sigma)\right|_{\sigma=0}\ .
\end{aligned}
\end{equation}
Expanding the exponential one finally obtains the bosonized fermion theory:
\begin{equation}\label{bosonization2}
\begin{aligned}
&\int\prod_{a=1}^N\prod_{t=1}^kd\eta_a^td\bar\eta_a^t~f\left(\sum_{a=1}^N\bar\eta_a^1\eta_a^1,\sum_{a=1}^N\bar\eta_a^1\eta_a^2,\ldots,\sum_{a=1}^N\bar\eta_a^k\eta_a^k\right)\\
&=\frac{1}{c_N^k}\int dU~e^{-N\text{~tr}\ln M}f(M_{11},M_{12},\ldots,M_{kk})\ ,
\end{aligned}
\end{equation}
where the integration is over the unitary part of the polar decomposition of the matrix $M$. The proof of the identity (\ref{det_J}) is given in the appendix of the reference \cite{kawamoto_smit}. However, the basic idea can be understood in the case $N=1$. Indeed, setting $R=1$ and factoring $U=e^{i\varphi}V$, where $V$ is a $SU(k)$ matrix one has:
\begin{equation}
\int dU\frac{e^{\text{~tr}JU}}{\det U}=\frac{1}{2\pi}\int_0^{2\pi}d\varphi\int_{SU(k)}dV~e^{-ik\varphi}~e^{e^{i\varphi}\text{~tr}JV}\ .
\end{equation}
Expanding the second exponential and integrating over $\varphi$ only the power of order $k$ gives a nonvanishing result. Therefore, one gets:
\begin{eqnarray*}
\int dU\frac{e^{\text{~tr}JU}}{\det U}&=&\frac{1}{k!}\int_{SU(k)}dV\left(\text{tr}JV\right)^k\\
&=&\frac{1}{k!}\sum_{t_1,\ldots,t_k}\sum_{t'_1,\ldots,t'_k}J_{t_1t'_1}\cdots J_{t_kt'_k}\int_{SU(k)}dV~V_{t'_1t_1}\cdots V_{t'_kt_k}\\
&=&\frac{1}{k!}\det J\ ,
\end{eqnarray*}
where we have used the group integration formula:
\begin{equation}
\int dV~V_{t'_1t_1}\cdots V_{t'_kt_k}=\frac{1}{k!}~\epsilon_{t'_1\ldots,t'_k}~\epsilon_{t_1\ldots,t_k}\ .
\end{equation}
This identity may be proven by generalizing the formula (\ref{det_U}) of the appendix \ref{group_integration_rules}.

Using the bosonized fermion theory (\ref{bosonization2}) one can show that the large $N$ solution of the fermion theory (\ref{fermion_theory}) is given at the saddle point $v\delta_{tt'}$ with:
\begin{equation}
v=\left.\frac{-m_f\left(1-\frac{1}{d}\right)+\sqrt{m_f^2+2d-1}}{d\left(1+\frac{m_f^2}{d^2}\right)}\right|_{m_f=0}=\frac{\sqrt{2d-1}}{d}\ ,
\end{equation}
whereas the free energy of the theory is:
\begin{equation}
{\cal F}=NN_tV\left[-\ln v+m_fv+dF(-\kappa^2v^2)\right]\ .
\end{equation}
This way, the chiral condensate is given by the expression:
\begin{equation}
\zeta=NN_t\sqrt{\frac{2}{d}}\sqrt{1-\frac{1}{2d}}\ .
\end{equation}
Therefore, whichever bosonization method is used one gets identical results.

Finally, let us compare these results with ours in the limit limit $m_f\to 0$. Our saddle point solution, equation (\ref{G1_o}), evaluated at $\kappa=1/2$, as well as the condensate, equation (\ref{condensate1}), read:
\begin{equation}
\left.G_o\right|_{m_f\to 0}=\frac{\sqrt{2d}}{d}\ ,~~~~~~~~\zeta=NN_t\sqrt{\frac{2}{d}}\ .
\end{equation}
Therefore, our approximation misses the prefactor $\sqrt{1-\frac{1}{2d}}$, which is of the order $O(1)$ even at $d=2$. For large $d$, the exact treatment of the function $F$, equation (\ref{F_function1}), yields the same result as ours.

\end{document}